\documentclass[journal]{IEEEtran}

\usepackage{cite}
\usepackage{amssymb,amsmath}
\usepackage{mathtools}
\usepackage{amsfonts, color}
\usepackage{url}
\usepackage{algorithm}
\usepackage{algpseudocode}
\usepackage{graphics,graphicx}
\usepackage{epstopdf , epsfig}
\usepackage{psfrag}
\usepackage{xcolor}
\usepackage{tikz}
\usepackage{pgfplots}\pgfplotsset{compat=newest}
\usepgfplotslibrary{fillbetween}
\usetikzlibrary{patterns}
\usetikzlibrary{shapes,arrows,shadows,positioning,decorations.pathreplacing,decorations.text,trees,calc}
\usepackage{pgfplots}
\usepackage[caption=false,font=footnotesize]{subfig}
\DeclareGraphicsExtensions{.eps}

\newtheorem{theorem}{Theorem}
\newtheorem{lemma}{Lemma}

\newtheorem{corollary}{Corollary}
\newtheorem{definition}{Definition}

\usepackage{latexsym}
\usepackage{slashed}
\usepackage{centernot}

\newcommand{\bt}{\mathbf}

\begin{document}
%
% paper title
% Titles are generally capitalized except for words such as a, an, and, as,
% at, but, by, for, in, nor, of, on, or, the, to and up, which are usually
% not capitalized unless they are the first or last word of the title.
% Linebreaks \\ can be used within to get better formatting as desired.
% Do not put math or special symbols in the title.
%\title{Rate-constrained Multiple Antenna Pooling for Interference Mitigation and Signal Enhancement}

\title{Interference Mitigation via Relaying}
%
%
% author names and IEEE memberships
% note positions of commas and nonbreaking spaces ( ~ ) LaTeX will not break
% a structure at a ~ so this keeps an author's name from being broken across
% two lines.
% use \thanks{} to gain access to the first footnote area
% a separate \thanks must be used for each paragraph as LaTeX2e's \thanks
% was not built to handle multiple paragraphs
%

%\author{S. Arvin~Ayoughi,~\IEEEmembership{Student Member,~IEEE,}
%        Wei~Yu,~\IEEEmembership{Fellow,~IEEE}% <-this % stops a space

\author{S. Arvin~Ayoughi, Wei~Yu,~\IEEEmembership{Fellow,~IEEE}

%\IEEEauthorblockA{
% \IEEEauthorrefmark{0}Department of Electrical and Computer Engineering \\
%    University of Toronto, Toronto, ON, M5S 3G4, Canada\\
%    Emails: sa.ayoughi@mail.utoronto.ca, weiyu@comm.utoronto.ca}}
\thanks{This work was supported in part by Huawei Technologies Canada Co., Ltd., and in part by the Natural Sciences and Engineering Research Council (NSERC) of Canada.
The materials in this paper have been presented in part at the IEEE International
Conference on Communication (ICC), London, UK, June 2015, and in part at the
$14^\mathrm{th}$ Canadian Workshop on Information Theory (CWIT), St. John's,
NL, Canada, July 2015.

The authors are with The Edward S. Rogers Sr. Department of
Electrical and Computer Engineering, University of Toronto, Toronto, Ontario M5S 3G4, Canada (e-mails: sa.ayoughi@mail.utoronto.ca, weiyu@comm.utoronto.ca).}}
\maketitle

% As a general rule, do not put math, special symbols or citations
% in the abstract or keywords.
\begin{abstract}
This paper studies the effectiveness of relaying for interference mitigation in an interference-limited communication scenario. We are motivated by the observation that in a cellular network, a relay node placed at the cell edge observes a combination of intended signal and inter-cell interference that is correlated with the received signal at a nearby destination, so a relaying link can effectively allow the antennas at the relay and at the destination to be pooled together for both signal enhancement and interference mitigation. We model this scenario by a multiple-input multiple-output (MIMO) Gaussian
relay channel with a digital relay-to-destination link of finite capacity, and with
\emph{correlated} noise across the relay and destination antennas. Assuming a
compress-and-forward strategy with Gaussian input distribution and quantization
noise, we propose a coordinate ascent algorithm for obtaining a stationary point of the non-convex joint optimization of the transmit and quantization covariance matrices. For fixed input distribution, the globally optimum quantization noise
covariance matrix can be found in closed-form using a transformation of the
relay's observation that simultaneously diagonalizes two conditional covariance
matrices by congruence. For fixed quantization, the globally optimum transmit
covariance matrix can be found via convex optimization. This paper further shows that
such an optimized achievable rate is within a constant additive gap of the MIMO relay channel capacity. The optimal structure of the quantization noise covariance enables a characterization of the slope of the
achievable rate as a function of the relay-to-destination link capacity. Moreover, this paper shows that the improvement in spatial degrees of freedom by
MIMO relaying in the presence of noise correlation is related to the aforementioned slope via a connection to the deterministic relay channel.  
\end{abstract}

% Note that keywords are not normally used for peerreview papers.

% For peer review papers, you can put extra information on the cover
% page as needed:
% \ifCLASSOPTIONpeerreview
% \begin{center} \bfseries EDICS Category: 3-BBND \end{center}
% \fi
%
% For peerreview papers, this IEEEtran command inserts a page break and
% creates the second title. It will be ignored for other modes.
\IEEEpeerreviewmaketitle

\section{Introduction}
\label{section_intro}

% The very first letter is a 2 line initial drop letter followed
% by the rest of the first word in caps.
% 
% form to use if the first word consists of a single letter:
% \IEEEPARstart{A}{demo} file is ....
% 
% form to use if you need the single drop letter followed by
% normal text (unknown if ever used by the IEEE):
% \IEEEPARstart{A}{}demo file is ....
% 
% Some journals put the first two words in caps:
% \IEEEPARstart{T}{his demo} file is ....
% 
% Here we have the typical use of a "T" for an initial drop letter
% and "HIS" in caps to complete the first word.

%\IEEEPARstart{C}{ooperation} among nodes that are equipped with large numbers of antennas enable the wireless network architecture to cope with the anticipated growth of demand for high data rates in next generations of this technology. Efficient use of capacity-limited cooperation links calls for optimized relaying schemes.

%\IEEEPARstart{C}{ooperative} communication and equipping devices with large number of antennas enable the next generations of wireless networks' architectures to cope with the anticipated growth of demand for high data rates \cite{fundamentalcoop,noncoop,newlook}. The main restrictive factor for increasing throughput in wireless networks is the interference. In a network, introducing cooperation at transmitters side or among receivers can effectively alleviate the interference, e.g.,\cite{newlook,txcoop,rxcoop}.

\IEEEPARstart{I}{nterference} is a main limiting factor for increasing data rates in many communication scenarios. A wireless cellular network with densely deployed basestations (BSs), for example, is typically interference limited. Provisioning of high data rates at cell edges, where signal is relatively weaker and interference is stronger, is a major challenge in cellular network physical layer design.  

%Interference is a main restrictive factor for increasing throughput in many wireless communication scenarios. Cellular networks with densely deployed base stations (BSs), for example, are typically highly interference-limited. Provisioning of high data rates at borders of cells, where signal is relatively weaker and interference is stronger, is a major challenge in physical layer design of these networks.

This paper explores the use of relays for interference mitigation.
The idea is that by placing a multiple-antenna relay at the cell edge
(see Fig.~\ref{system}), a user device in close proximity of a relay
would be able to establish an out-of-band relaying link and benefit from
relay's observation in decoding the downlink signal. Due to the physical proximity of the relay and the intended receiver, their observed interference signals are highly correlated. In essence, the relay and the intended receiver would be able to pool their antennas together using the relaying link. The extra spatial dimensions enables not only signal enhancement but also interference mitigation at the receiver.

%When treated as noise, interference can also be mitigated if a dedicated relay device provides receivers with extra spatial dimensions \cite{universal,incremental}.

The benefit of antenna pooling depends crucially on the quality of
the relaying link. The goal of this paper is to quantify the benefit of
relaying for downlink transmission of a cellular network as a function of the relaying link capacity. 
Toward this end, this paper studies a multiple-input multiple-output
(MIMO) Gaussian relay channel with digital relaying link, 
%and $s$, $d$, $r$, and $t$ antennas at source, destination, relay, and the interference source, respectively. 
where, due to common inter-cell interference, the noise processes
across the relay and destination antennas are correlated. This is a
simple yet fundamental model of cooperative communications for which
information theoretical analysis can yield significant insight into the effectiveness of cooperative interference mitigation. 
In particular, this paper adopts a compress-and-forward relaying
strategy, which is appropriate given the physical proximity of the
relay and destination in the scenario of interest \cite{caprelay,
kgg}. In this case, the relay provides the destination with a
compressed version of its observations, the accuracy of which is
determined by the available capacity of the relaying link.
The relay uses Wyner-Ziv coding to exploit the side information
available at the destination in quantizing its observations. The goal of this
paper is to analyze the optimal transmission and quantization
strategies for the MIMO relay channel with digital relaying link in
the presence of correlated interference.

\subsection{Main Results}
\label{section_main}

%The joint optimization of the input transmission and the compression process is crucial for characterizing the compress-and-forward rate for the MIMO relay channel. Toward the goal of understanding how to best take full advantage of the MIMO relay, 
This paper makes the following contributions toward the goal of
understanding how to best take advantage of the MIMO relay for
both signal enhancement and interference mitigation:

\begin{itemize}
\item We propose an iterative algorithm for optimizing the
covariance matrices of Gaussian input signal and Gaussian quantization 
noise for the MIMO relay channel with correlated noise. 
%we propose an efficient iterative algorithm. The approach is to maximize the Lagrangian for a fix Lagrange multiplier by coordinate ascent. The obtained solution is, then, used to find the optimal Lagrange multiplier in an outer loop. Despite non-convexity of the problem, the algorithm converges to a stationary point.
We show that the optimization of input covariance matrix for fixed
quantization noise distribution is a concave optimization, and the
optimization of quantization noise covariance matrix for fixed input
distribution can be solved via a simultaneous diagonalization
transformation.
Further, the allocation of relaying bits across the spatial dimensions
should follow a reverse water-filling solution, where 
%transformed into a convex problem. Here, optimal quantization scheme is to transform the relay's observed vector using a particular matrix, followed by a reverse water-filling solution to allocate relay-destination link's capacity for describing each element.
more bits are used to quantize spatial dimensions with higher
conditional signal-to-interference-and-noise ratios (CSINRs). 
%An element with zero CSINR is never quantized. We show that at most $\min(r, s)$ elements should be quantized. Furthermore, we characterize conditions on the numbers of antennas under which the CSINR goes to infinity as background noise goes to zero.

\item We characterize the slope of compress-and-forward achievable rate versus the relaying link capacity curve for the optimized quantization noise
covariance at a fixed input distribution. The slope is related to the
generalized eigenvalues of certain conditional covariance matrices. One of the main results of this paper is that this slope can asymptotically approach the maximum value of 1, if the MIMO relay channel contains an asymptotically deterministic component.

\item We characterize the improvement in the spatial degrees of freedom (DoF) by
compress-and-forward relaying as a function of the number of antennas in the
network. A distributed interference zero-forcing scheme at the relay is shown
to be DoF optimal. The improvement in spatial DoF (assuming infinite relay link
capacity) is shown to be equal to the number of asymptotically deterministic
component in the channel. Thus, the existence of asymptotically deterministic
components is the fundamental reason for the effectiveness of relaying for
interference mitigation and signal enhancement.
%Appropriate transformation at the relay before quantization saves DoF of the relay-destination link.

%\item We show that, at high signal-to-noise-ratio (SNR) and
%interference-to-noise-ratio (INR), the slope of compress-and-forward rate curve versus
%relaying link capacity approaches its maximum value of 1,
%if and only if relaying improves the overall DoF.  

\item The optimized compress-and-forward strategy is shown to achieve the
capacity of the MIMO relay channel with noise correlation to within a constant additive gap, which only depends on the number of antennas.% at transmitter and the relay.  % of $\min(r, s)$ bits per channel use.

\end{itemize}

\subsection{Related Works}
\label{section_relwrk}

The relay channel is a classical model that has been widely studied in
the literature. Specifically related to this work, relaying in the presence of noise correlation for single-input
single-output (SISO) Gaussian channel is studied in \cite{corr}, where
the negative correlation between relay and destination noises is
shown to improve the achievable rate of the compress-and-forward
relaying scheme. For the Gaussian MIMO relay channel with independent
noises, a coordinate ascent procedure for maximizing the compress-and-forward
achievable rate over input and quantization noise covariance matrices
is proposed in \cite{del}. %In this problem, the relay quantization rate constraint depends on both optimization variables. Therefore, the convergence of coordinate ascent cannot be guaranteed if, as in \cite{del}, in each iteration the constraint is satisfied with equality \cite{bertsekas}. To guarantee convergence to a Karush-Kuhn-Tucker (KKT) solution, in this paper we propose alternating optimization of Lagrangian.
For optimizing quantization at the relay under fixed input
distribution, \cite{del} uses the Conditional Karhunen-Lo\`{e}ve
Transform (CKLT) \cite{klt} of the relay's observed vector given the destination's observation, followed by a reverse water-filling solution for quantization rate allocation. %After transformation, elements of the relay's observed vector are quantized independently. Allocating the relaying capacity for quantizing each element is related to the eigenvalues of a certain conditional covariance matrix with a reverse water-filling interpretation. In this paper, we focuses on the MIMO
This paper focuses on the relay channel with correlated noises for
which the solution is more complex. We show that the right
transformation is the simultaneous diagonalization by
*congruence\footnote{Here, ``*" refers to conjugate transpose of the
transformation matrix. Although in this paper we use $(.)^\dagger$
to denote conjugate transpose, we preferred not to change the
terminology of \cite{horn}.}
\cite{horn} of two conditional covariance matrices of relay observation \cite{icc}. 
%After this transformation, the elements of the vector become conditionally independent given destination observation as well as given both destination observation and input signal. 
The optimal allocation of quantization rates again has a reverse
water-filling interpretation, and is related to the generalized
eigenvalues of the conditional covariance matrices. When noises 
are independent simultaneous diagonalization transform simplifies to the CKLT. 

The optimization of quantization noise covariance is also solved in
\cite{JunChen}, but from a source coding perspective. In \cite{JunChen}, canonical correlation analysis (CCA) is used to transform the relay observation before quantization. CCA can be interpreted as indirect simultaneous diagonalization by *congruence of
two conditional covariance matrices of the relay observation.
Although our solution to the optimization problem can eventually be
shown to be the same as that of \cite{JunChen}, the direct
diagonalization approach in this paper is simpler and provides insight
into the optimized MIMO relaying strategy for interference mitigation
and signal enhancement. 

For the SISO Gaussian relay channel with noise correlation, \cite{lei}
shows that compress-and-forward achieves the capacity to within a
constant additive gap. For the MIMO Gaussian relay channel with
independent noise vectors, capacity approximation using the partial
decode-and-forward scheme is provided in \cite{appcap} using partial decode-and-forward scheme. In this paper, for the MIMO Gaussian relay channel with noise
correlation, we use the simultaneous diagonalization transform to show
that compress-and-forward achieves the capacity to within a constant
additive gap that is tighter than the gap of extension of \cite{lei}
to the MIMO case.

This paper studies the improvement in spatial DoF due to the
compress-and-forward relaying. %Quantizing elements of relay observation vector independently without any transformation wastes the prelog factor of the relay-destination link capacity. 
The DoF-optimal transformation at the relay prior to quantization
involves distributed zero-forcing of interference. Distributed
zero-forcing of interference has been considered for various classes
of SISO relay networks with analog relaying links, e.g., \cite{neut,
ain,rankov,jafarcorr}. This paper deals with a MIMO relay channel with
digital relaying link in which distributed zero-forcing of
interference reveals the asymptotically deterministic components of
the MIMO relay channel. The determinism here refers to the condition
that the observation of the relay is a deterministic function of the
input of the channel and the observation of the destination. As shown in
\cite{coverkim}, compress-and-forward achieves the cut-set upper bound in
this case. This paper illustrates that this type of determinism occurs
in the MIMO relay channel in the asymptotically high SNR and INR regime. This is the fundamental reason that compress-and-forward relaying can be effective in improving the overall throughput in a MIMO relay channel \cite{cwit}.

%In \cite{neut}, for a class of Gaussian relay-interference networks 
%introduced interference neutralization 
%For the single-input single-output (SISO) Gaussian $2 \times 2 \times 2$-IC interference neutralization achieves approximately optimal rates in partially connected scenarios by enabling over-the-air interference cancellation through analog relay-destination links \cite{neut}.

%The optimal DoF of the $2 \times 2 \times 2$-IC is achieved by a careful design of beamforming vectors at transmitters to have alignment at the relays and combining vectors at the relays for distributed zero-forcing of interference at destinations, called aligned-interference-neutralization \cite{ain}.

%In \cite{rankov} distributed zero-forcing is used for improving spectral efficiency of half-duplex two-hop relay networks.

%Amplify-and-forward over SISO relay networks in studied in \cite{jafarcorr} and the optimal relay amplification vector is derived. Furthermore, it is shown that in a network relays can perform distributed interference cancellation if the number of relays is greater than the number of interferers \cite{jafarcorr}.

%\begin{figure}
%\center
%\def\svgwidth{0.4\textwidth}
%\input{system9.eps_tex}
%\small
%\caption{A cellular network with a BS and two cell-edge devices that pool antennas through a finite-capacity digital link. The relay and the destination devices experience correlated noise due to common interference signals.}
%\label{system}
%\end{figure}

\begin{figure}
\center
\def\svgwidth{0.4\textwidth}
\begingroup%
  \makeatletter%
  \providecommand\color[2][]{%
    \errmessage{(Inkscape) Color is used for the text in Inkscape, but the package 'color.sty' is not loaded}%
    \renewcommand\color[2][]{}%
  }%
  \providecommand\transparent[1]{%
    \errmessage{(Inkscape) Transparency is used (non-zero) for the text in Inkscape, but the package 'transparent.sty' is not loaded}%
    \renewcommand\transparent[1]{}%
  }%
  \providecommand\rotatebox[2]{#2}%
  \ifx\svgwidth\undefined%
    \setlength{\unitlength}{884.11125728bp}%
    \ifx\svgscale\undefined%
      \relax%
    \else%
      \setlength{\unitlength}{\unitlength * \real{\svgscale}}%
    \fi%
  \else%
    \setlength{\unitlength}{\svgwidth}%
  \fi%
  \global\let\svgwidth\undefined%
  \global\let\svgscale\undefined%
  \makeatother%
  \begin{picture}(1,0.74908559)%
    \put(0,0){\includegraphics[width=\unitlength]{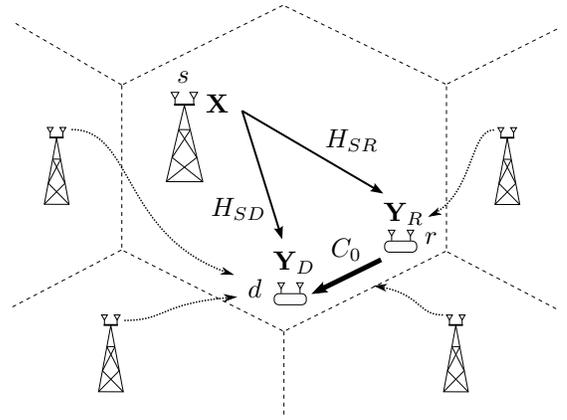}}%
    \put(0.5841406,0.29037362){\color[rgb]{0,0,0}\makebox(0,0)[lb]{\smash{$C_0$}}}%
    \put(0.35488155,0.55439956){\color[rgb]{0,0,0}\makebox(0,0)[lb]{\smash{$ \bt X$}}}%
    \put(0.68116382,0.35546042){\color[rgb]{0,0,0}\makebox(0,0)[lb]{\smash{$ \bt Y_R$}}}%
    \put(0.478341639,0.26595671){\color[rgb]{0,0,0}\makebox(0,0)[lb]{\smash{$ \bt Y_D$}}}%
    \put(0.57388117,0.49115255){\color[rgb]{0,0,0}\makebox(0,0)[lb]{\smash{$ H_{SR}$}}}%
    \put(0.36445558,0.36556765){\color[rgb]{0,0,0}\makebox(0,0)[lb]{\smash{$H_{SD}$}}}%
    \put(0.3024066307,0.60908359){\color[rgb]{0,0,0}\makebox(0,0)[lb]{\smash{$s$}}}%
    \put(0.43243302,0.2163403){\color[rgb]{0,0,0}\makebox(0,0)[lb]{\smash{$d$}}}%
    \put(0.756417796,0.31354995){\color[rgb]{0,0,0}\makebox(0,0)[lb]{\smash{$r$}}}%
  \end{picture}%
\endgroup%
\small
\caption{A cellular network with a BS and a cell-edge user that is located close to a relay node. The relay pools antennas with the receiver through a finite-capacity digital link. Due to the common interference, the noise processes at the relay and at the destination antennas are correlated.}
\label{system}
\end{figure}

\subsection{Notation}

In this paper, we denote matrices by uppercase letters, e.g., $H$,
vectors of random variables by uppercase bold letters, e.g., $\bt X$,
and its realization by lowercase bold letters, e.g., $\bt x$. Also,
$\bt 0_{r \times 1}$ stands for the $r$-dimensional zero vector and
$\bt I_r$ is the $r \times r$ identity matrix. Conjugate transpose,
Moore-Penrose pseudoinverse, trace, determinant, rank, the
$i^\mathrm{th}$ largest eigenvalue, and the row span of a matrix are
denoted by $\left(.\right)^\dagger$, $\left(.\right)^{-1}$,
$\mathrm{tr}\left(.\right)$, $\left|.\right|$,
$\mathrm{rank}\left(.\right)$, $\lambda_i\left(.\right)$, and
$\mathrm{rowspan}\left(.\right)$, respectively. For two matrices $A$ and $B$, $A \succeq B$ means that $A-B$ is positive semidefinite. The expectation
operator is shown by $\mathbb{E}(.)$. The set of complex numbers is
denoted by $\mathbb{C}$. Finally, we define $\left( x\right)^+ =
\max(0,x)$.

\subsection{Organization of the Paper}

Section II introduces the system model, relaying scheme, and the
problem formulations. The proposed iterative algorithm for optimizing
the compress-and-forward scheme is presented in Section III.
Section IV provides interpretations for the optimal structure of
quantization noise and characterizes the sensitivity of the achievable rate to changes in the relaying link capacity. Section V provides a DoF analysis of the MIMO relay channel.
Section VI interprets the DoF analysis via a connection to deterministic
relay channel. Section VII shows that compress-and-forward achieves 
the capacity of the MIMO Gaussian relay channel with noise correlation 
to within a constant additive gap. Simulation results are 
presented in Section VIII, and Section IX concludes the paper.

\section{MIMO Gaussian Relay Channel with Noise Correlation}
\label{section_mimo_relay}
\subsection{System Model}
\label{section_sys_model}

Consider the transmission from a BS to a cell-edge user that is located in close proximity of a relay node in a wireless cellular network, as shown in
Fig.~\ref{system}. The relay and the destination observe common interference from nearby BSs and treat it as noise. Due to the common interference, the observed noises across the relay and the destination antennas are correlated. An out-of-band relay link is established between the relay and the user device for interference mitigation and signal
enhancement. 

Mathematically, this communication scenario is modeled as a Gaussian
MIMO relay channel with noise correlation and with a digital relaying link of capacity $C_0$ bits. The source, relay, and destination are equipped with $s$, $r$, and $d$ antennas, respectively. Let $t$ be the total number of antennas from
all the interfering BSs. The received signals at the relay
and the destination can be written as
\begin{align}
	\bt Y_R &= H_{SR}\bt X + \bt N_R, \label{sys_model_1} \\
	\bt Y_D &= H_{SD}\bt X +\bt N_D,
\label{sys_model_2}
\end{align}
where
\begin{align}
	\bt N_R &= H_{TR}\bt X_T +\bt N_1, \label{noise_1} \\
	\bt N_D &= H_{TD}\bt X_T +\bt N_2
\label{noise_2}
\end{align}
are the correlated noise vectors. Here, $ H_{SR} \in \mathbb{C}^{r \times s}$
and $ H_{SD} \in \mathbb{C}^{d \times s}$ are the source-relay and
source-destination channel matrices respectively; $ H_{TR} \in
\mathbb{C}^{r \times t}$ and $ H_{TD} \in \mathbb{C}^{d \times t}$ are
the interferers-to-relay and interferers-to-destination
channel matrices respectively; 
$\bt N_1 \sim \mathcal {CN} (\bt 0_{r \times 1}, \sigma ^2\bt I_r )$
and $\bt N_2 \sim \mathcal {CN} (\bt 0_{d \times 1}, \sigma ^2\bt I_d
)$ are the additive and independent background noises at the relay
and the destination respectively; $\bt X \in \mathbb{C}^{s \times 1}$
is the transmit signal vector from the source under power constraint
$\mathbb{E}\left(\bt X^\dagger \bt X\right) \leq P$; finally $\bt X_T
\in \mathbb{C}^{t \times 1}$ is the interference signal vector that is
assumed to be Gaussian with $\bt X_T \sim \mathcal {CN} \left(\bt 0_{t
\times 1}, S_{ \bt X_T}\right)$, independent of everything else, and
is treated as a part of noise. In order to carry
out DoF analysis, we assume that
the entries of the channel matrices are drawn independently from a
continuous probability distribution, so that each of them as well
as their concatenations are full-rank almost surely. Throughout this paper, we assume that the channel state information is known perfectly.

\subsection{Capacity Upper and Lower Bounds}
\label{section_upperbound}

In this section, we state the cut-set upper bound as well as two expressions for the compress-and-forward lower bound on the capacity of the relay channel. We use these well-known results through out the paper.

The capacity of the relay channel is upper bounded by {\cite[Theorem 4]{caprelay}}:
\begin{align}
C \leq \max_{\substack{p(\bt x), \\ \mathbb E \{ \bt X^{\dagger} \bt X \} \leq P}} \min \{  I( \bt X ; \bt Y_R  , \bt Y_D ) , I( \bt X ; \bt Y_D ) + C_0 \}.
\label{csb_max_min}
\end{align}
The evaluation of this bound, known as cut-set upper bound, is a convex optimization
problem. Here, the optimal $p(\bt x)$ is a multivariate Gaussian, i.e., 
$\bt X \sim \mathcal {CN} \left(\bt 0_{s \times 1}, S_{\bt X}\right)$ 
for some positive semidefinite $S_{\bt X}$. %The two mutual information terms in cut-set bound are both concave functions of $S_{\bt X}$.

%the optimization of the covariance
%matrix $S_{\bt X}$ in (\ref{csb_max_min}) can be written as
%\begin{equation}
%\begin{array}{llcl}
%C & \leq &\displaystyle \max_{S_{ \bt X}, z} & z \\ \\
%&& \mathrm{s.t.} &  z \leq I( \bt X ; \bt Y_r  , \bt Y_d ), \\ \\
%&&& z \leq I( \bt X ; \bt Y_d ) + C_0, \\ \\
%&&& \mathrm{tr}(S_{\bt X}) \leq P, \\ \\
%&&& S_{\bt X} \succeq \bt 0,
%\end{array}
%\label{csb_max}
%\end{equation}
%which is a convex problem.
%
%
%\subsection{Relaying Scheme}
%\label{section_lowerbounds}

%To enable antenna pooling for cooperative interference mitigation and signal enhancement, in 

%The relay node compresses its observations by quantizing it and forwards the quantized version to the destination.  Due to both transmission from source and correlation of noises the relay's and destination's observations are statistically correlated.  Wyner-Ziv coding exploits this correlation in quantizing relay's observation to use the finite capacity of relay-destination link more efficiently. The achievable rate of this scheme is given in the following theorem.

%In decoding scheme of compress-and-forward in \cite{caprelay} the quantized version of the relay's observation is uniquely decoded. The message is then decoded from destination's observation and the quantized version of the relay's observation. Decoding this quantized version uniquely requires the quantization rate at the relay to not to exceed the relay-destination capacity. The achievable rate is given in the following theorem.

The capacity of the relay channel is lower bounded by {\cite[Theorem 6]{caprelay}}:
\begin{equation}
\begin{array}{llcl}
C & \geq &\displaystyle \max_{ p(\bt x)p(\hat{\bt y}_R \vert \bt y_R)} & I( \bt X ; \widehat{\bt Y}_R  , \bt Y_D ) \\
&& \mathrm{subject~to} &  I( \bt Y_R ; \widehat{ \bt Y }_R \vert \bt Y_D ) \leq C_0, \\
&&& \mathbb E \{ \bt X^{\dagger} \bt X \} \leq P.
\end{array}
\label{cf1}
\end{equation}
%In decoding scheme of \cite{minenergy} for compress-and-forward, however, the message is decoded by non-unique decoding of the quantized version of the relay's observation. The resulting achievable rate is given in the next theorem.
%\begin{equation}
%\begin{array}{rrcll}
%R_{CF,2} &=& \displaystyle \max_{\substack{p(\bt x)p(\hat{\bt y}_r \vert \bt y_r), \\ \mathbb E \{ \bt X^{\dagger} \bt X \} \leq 1}} & \min \left\{ I( \bt X ; \widehat{\bt Y}_r  , \bt Y_d ), \right.  \\
%&&& I( \bt X ; \bt Y_d ) + C_0 - \\
%&&& \left. I( \bt Y_r ; \widehat{ \bt Y }_r \vert \bt Y_d, \bt X ) \right\}
%\end{array}
%\end{equation}
The evaluation of this bound, known as the compress-and-forward rate, is not
always straightforward. For the Gaussian relay channel, although it
can be shown that Gaussian quantization at the relay is optimal for
Gaussian signaling at the source \cite{JunChen} and vice versa, the jointly optimal input distribution and quantization test channel of the compress-and-forward lower bound is not yet known; see \cite{diamond} for an example where jointly Gaussian distribution is suboptimal. For tractability, this paper restricts attention to jointly Gaussian transmission $\bt X \sim \mathcal{CN} 
(\bt 0_{s \times 1}, S_{ \bt X})$ and Gaussian quantization modeled as 
\begin{align}
\widehat{\bt Y}_R = \bt Y_R + \bt Q,
\label{testchannel}
\end{align}
where the quantization noise $\bt Q \sim \mathcal{CN} (\bt 0_{r
\times 1},S_{\bt Q})$ is independent of all variables. Even then, the
optimization of the achievable rate over $(S_{\bt X}, S_{\bt Q})$ is
still not straightforward. This optimization is a main subject of this
paper.

In the above compress-and-forward scheme, the destination first
decodes the quantized version of the relay's observation uniquely,
then proceeds to decode the source message. Reliable unique decoding of the
quantization codeword requires the compression rate at the relay not
to exceed the relaying link capacity. In an alternative decoding
scheme, the destination decodes the message by non-unique decoding of the
quantization codeword at the relay, thus allowing the compression rate
at the relay to potentially exceed the relay link capacity. Although
it can be shown that such a flexibility does not result in a higher
achievable rate, the resulting rate expression has the advantage of resembling the cut-set bound,
% in (\ref{csb_max_min}). Also, by removing the relay-destination link capacity constraint, it allows evaluation at an arbitrary input distribution and quantization test channel. 
which is useful in characterizing the capacity of the relay channel to within a constant gap. This alternative rate expression is {\cite[Theorem 16.4]{nit}}:
\begin{multline}
C \ \geq  \displaystyle \max_{\substack{p(\bt x)p(\hat{\bt y}_R \vert \bt y_R), \\ \mathbb E \{ \bt X^{\dagger} \bt X \} \leq P}}  \min \left\{  I( \bt X ; \widehat{\bt Y}_R, \bt Y_D ), \right.  \\  \left.   I( \bt X ; \bt Y_D ) + C_0 - I( \bt Y_R ; \widehat{ \bt Y }_R \vert \bt Y_D, \bt X ) \right\}.
\label{cf2}
\end{multline}
Optimization problems (\ref{cf1}) and (\ref{cf2}) have
the same global maximum, and their optimal distributions $p(\bt
x)p(\hat{\bt y}_R \vert \bt y_R)$ are equal.

\subsection{Problem Formulation}
\label{section_problem}

This paper addresses four aspects of MIMO compress-and-forward
relaying in the presence of noise correlation. 

\subsubsection{Optimization of Input and Quantization Covariance
Matrices}
We propose a method for optimizing the achievable rate (\ref{cf1}). Assuming
jointly Gaussian input distribution and quantization test channel, the
achievable rate (\ref{cf1}) can be expressed as

%\begin{subequations} 
\begin{equation}
\begin{array}{llcl}
R_{CF}(C_0) & = &\displaystyle \max_{S_{ \bt X}, S_{\bt Q}} & f_o(S_{\bt X}, S_{\bt Q}) \\
&& \mathrm{subject~to} &  f_c(S_{\bt X}, S_{\bt Q}) \leq C_0,  \\
&&& \mathrm{tr}( S_{\bt X}) \leq P,  \\
&&& S_{ \bt X} \succeq \bt 0, \  S_{\bt Q}  \succeq \bt 0,
\end{array}
\label{opt_prob}
\end{equation}
%\end{subequations}
where the objective function is
\begin{multline}
f_o (S_{\bt X},S_{\bt Q}) = I( \bt X ; \widehat{ \bt Y}_R, \bt Y_D )  \\ = \log \frac{ \left|  H S_{\bt X}  H^{\dagger} + S_\bt{int} + \sigma^2 \bt I_{(r+d)} + 
\begin{bmatrix} S_{\bt Q} &  \bt 0_{r\times d} \\  
	\bt 0_{d \times r} & \bt 0_{d\times d} \end{bmatrix} \right|}{ \left|  S_\bt{int} + \sigma^2 \bt I_{(r+d)} + 
\begin{bmatrix} S_{\bt Q} &  \bt 0_{r\times d} \\  
	\bt 0_{d \times r} & \bt 0_{d\times d} \end{bmatrix} \right|},
\label{opt_objective}
\end{multline}
and the constraint is
\begin{multline}
f_c \left(S_{\bt X},S_{\bt Q}\right) = I(\bt Y_R ; \widehat{ \bt Y}_R \mid \bt Y_D) \\  = \log \frac{\left|  H S_{\bt X}  H^{\dagger} + S_\bt{int} + \sigma^2 \bt I_{(r+d)} + 
\begin{bmatrix} S_{\bt Q} &  \bt 0_{r\times d} \\  
	\bt 0_{d \times r} & \bt 0_{d\times d} \end{bmatrix} \right|}{ \left| H_{SD} S_{ \bt X}  H_{SD}^{\dagger} +
S_\bt{int}^{\scriptscriptstyle (2,2) }+\sigma ^2\bt I_d \right| \left| S_{\bt Q} \right|},
\label{opt_constraint}
\end{multline}
where $H = \big[ H_{SR}^{\dagger} \  H_{SD}^{\dagger} \big]^{\dagger}$, and
\begin{equation}
S_\bt{int} = 
\begin{bmatrix} S_\bt{int}^{\scriptscriptstyle (1,1)} &  S_\bt{int}^{\scriptscriptstyle (1,2)} \\  
	S_\bt{int}^{\scriptscriptstyle (2,1)} &  S_\bt{int}^{\scriptscriptstyle (2,2)} \end{bmatrix} \\  =
\begin{bmatrix} H_{TR} S_{ \bt X_t}  H_{TR}^{\dagger} &
H_{TR} S_{ \bt X_t}  H_{TD}^{\dagger} \\
H_{TD} S_{ \bt X_t}  H_{TR}^{\dagger} &
H_{TD} S_{ \bt X_t}  H_{TD}^{\dagger}
\label{S_int}
\end{bmatrix} 
\end{equation}
is the interference covariance matrix. %We assume channel state information and the statistics of noise and interference are available at a centralized location for solving optimization problem (\ref{opt_prob}).

\subsubsection{Characterization of the Slope of Achievable Rate with Respect to $C_0$} 

We evaluate the effectiveness of compress-and-forward in 
improving the overall rate in a relay channel as measured by 
the slope
\begin{align}
\frac{d \bar R_{CF}(C_0)}{d C_0},
\label{slope_def}
\end{align}
where $\bar R_{CF}(C_0)$ is the achievable rate expression (\ref{opt_prob}) evaluated at a fixed $S_{\bt X}$. We argue that $\bar R_{CF}(C_0)$ is concave in $C_0$. Therefore, its maximum slope occurs at $C_0 = 0$. By the upper bound (\ref{csb_max_min}), this slope cannot exceed 1. This paper provides conditions under which the slope is asymptotically
close to its maximum value of 1.

%We show that for larger CSINR of relay's observations given destination's observations the slope is closer to 1.

\subsubsection{DoF Improvement by Compress-and-Forward} 

We also evaluate the effectiveness of compress-and-forward in the
large $C_0$ and high SNR and INR regime by studying the improvement in spatial DoF due to relaying. In particular, define 
\begin{align}
\rho \triangleq \frac{1}{\sigma^2},
\label{snr-inr}
\end{align}
and let the relaying link capacity scale with $\rho$ as
\begin{align}
C_0(\rho) = \alpha \log\left(\rho \right) + o\left(\log\left(\rho \right) \right).
\label{c0}
\end{align}
The DoF improvement due to relaying is defined as
\begin{align}
\Delta DoF \triangleq \lim_{\rho \to \infty}
\frac{R_{CF}\left(C_0(\rho) \right)-R_{CF}(0)}{\log \left(\rho \right)}.
\label{dof-def}
\end{align}
We show that the conditions on the number of antennas in the system 
$\left(s, d, r, t\right)$ under which relaying with $\alpha = \infty$ brings in a DoF improvement are identical to the conditions on the number of antennas under which at small $C_0$ the slope (\ref{slope_def}) asymptotically approaches its maximum value.

% The prelog factor of the rate expression is the achieved number of independent and interference-free scaler channels.

\subsubsection{Characterizing the Capacity to Within a Constant Gap} 

Finally, we show the constant-gap optimality of the
compress-and-forward strategy for the MIMO relay channel with noise
correlation. We bound the gap between the achievable rate
(\ref{opt_prob}) and the cut-set bound (\ref{csb_max_min}) by a
constant that only depends on the number of antennas. %In the absence of an exact capacity characterization, the constant gap characterization provides an optimality guarantee, which is relevant in the high SNR regime.

\section{Optimization of Compress-and-Forward}
\label{section_algo}

The joint optimization of transmission at the source and quantization at 
the relay is crucial for efficient use of the relay. This section provides a method for joint optimization of $S_{\bt X}$ and $S_{\bt Q}$, and illustrates the structure of $S_{\bt X}$ and $S_{\bt Q}$ at the stationary point of the overall optimization problem
(\ref{opt_prob}). 

\subsection{Iterative Optimization of Lagrangian}
\label{section_lagrangian_max}

The joint optimization of the input and quantization covariance
matrices is not a convex optimization problem, as both the objective
function and the constraint of (\ref{opt_prob}) are concave in 
$S_{\bt X}$ and convex in $S_{\bt Q}$. Our approach for tackling (\ref{opt_prob}) is to first find a stationary point of the Lagrangian  
\begin{align}
\mathcal{L}( S_{\bt X}, S_{\bt Q}, \mu)= 
f_o (S_{\bt X},S_{\bt Q}) - \mu (f_c (S_{\bt X},S_{\bt Q}) - C_0),
\label{lagrangian}
\end{align}
for a fixed Lagrange multiplier $\mu$ by solving
\begin{equation}
\begin{array}{cl}
\underset{S_{ \bt X}, S_{\bt Q}}{\mathrm{maximize}} & \mathcal{L}( S_{\bt X},S_{\bt Q}, \mu) \\
\mathrm{subject~to} &  \mathrm{tr}( S_{\bt X}) \leq P,  \\
& S_{ \bt X} \succeq \bt 0, \  S_{\bt Q}  \succeq \bt 0,
\end{array}
\label{opt_lagrangian}
\end{equation}
then to search for the $\mu$ that results in
\begin{align} 
f_c (S_{\bt X}^*, S_{\bt Q}^*) = C_0,
\label{comp_slack}
\end{align}
in an outer loop. It can be easily shown that the optimal $\mu^* \in (0,1)$, 
%When $\mu=0$, i.e., when the objective is to maximize the overall rate without relay link capacity constraint, the optimal $S_{\bt Q}^*$ is a zero matrix, resulting in $f_c(S_{\bt X}^*, S_{\bt Q}^*) = +\infty$.
because if $\mu \ge 1$, i.e., if the relay link capacity constraint
penalizes the objective at more than a 1:1 ratio, then the optimal 
$S_{\bt Q}^*$ would be infinite, resulting in $f_c(S_{\bt X}, S_{\bt Q}^*) = 0$. 
Thus, finding $\mu^*$ is a one-dimensional root-finding problem. 
If $f_c(S^*_{\bt X}(\mu), S_{\bt Q}^*(\mu))$ is continuous in $\mu \in (0, 1)$, 
then the optimal $\mu$ that satisfies (\ref{comp_slack}) can be found by bisection. 
In case of discontinuity, time-sharing between the two operating points is needed.

We propose an iterative coordinate ascent approach for solving
(\ref{opt_lagrangian}). In an iteration of this coordinate ascent procedure, we obtain the global optimum of $S_{\bt X}$ and $S_{\bt Q}$ while keeping the other variable fixed (and the global optimum is essentially unique). This iteration process generates a nondecreasing sequence of the Lagrangian objective values; hence, it converges. %The idea is to find the optimal $S_{\bt Q}$ that maximizes $\mathcal{L}( S_{\bt X}, S_{\bt Q}, \mu)$ for a fixed $S_{\bt X}$, then to find the optimal $S_{\bt X}$ that maximizes $\mathcal{L}(S_{\bt X}, S_{\bt Q}, \mu)$ for a fixed $S_{\bt Q}$, and to iterate between the two steps. For a fixed $\mu \in \left(0,1\right)$, each of the individual optimizations of $S_{\bt X}$ and $S_{\bt Q}$ can be solved to global optimality (and the global optimum is essentially unique). The iterative optimization process generates a nondecreasing sequence of the Lagrangian objective, hence it converges.

\begin{algorithm}[t]
\caption{Joint Input and Quantization Optimization (\ref{opt_prob})}
\label{algo}
\begin{algorithmic}[1]
\State Initialize $S_{\bt X} \succeq \bt 0$ such that $\mathrm{tr}\left(S_{\bt X}\right) = P$;
\Repeat
\State For a fixed $\mu$:
\Repeat
\State {Find optimal $S_{\bt Q}$ for fixed $S_{\bt X}$ as in Section \ref{section_opt_S_Q};}
\State {Find optimal $S_{\bt X}$ for fixed $S_{\bt Q}$ as in Section \ref{section_opt_S_X};}
\Until {Convergence;}
\State {Update $\mu$ using bisection;} 
\Until {$f_c\left(S_{\bt X}, S_{\bt Q}\right)=C_0$.}
\end{algorithmic}
\end{algorithm}

Details of optimizing $S_{\bt Q}$ for a fixed $S_{\bt X}$ and
optimizing $S_{\bt X}$ for a fixed $S_{\bt Q}$ are provided in the
subsequent sections. The overall iterative approach is summarized as
Algorithm \ref{algo}.  The following theorem states the convergence
result formally.

\begin{theorem}
Assuming that the optimal $S_{\bt X}$ for a fixed $S_{\bt Q}$ is unique and 
the optimal $S_{\bt Q}$ for a fixed $S_{\bt X}$ is unique, the inner iterative
optimization procedure in Algorithm \ref{algo} converges to a stationary point
of the Lagrangian maximization problem (\ref{opt_lagrangian}). Further, if 
bisection finds the $\mu$ that satisfies (\ref{comp_slack}), then such a $\mu$ leads
to a Karush-Kuhn-Tucker (KKT) point of the joint transmit and quantization
noise covariance optimization problem (\ref{opt_prob}). %Assuming that the constraint function () is contentious and decreasing in $\mu$, 
\label{thm_algo}
\end{theorem}

\begin{IEEEproof} For a fixed $\mu$, coordinate ascent on the Lagrangian is monotonically increasing; hence, it is convergent. The uniqueness of solution in the optimization of $S_{\bt X}$ for a fixed $S_{\bt Q}$ and in the optimization of $S_{\bt Q}$ for a fixed $S_{\bt X}$ ensures that coordinate ascent converges to a stationary point.
This together with $\mu$ that satisfies (\ref{comp_slack}) gives a KKT point of (\ref{opt_prob}).
\end{IEEEproof}

There are special cases, for example when $s=1$ or $r=1$, where the above procedure would produce a globally optimal solution. However, in general only the convergence to a stationary point is assured.

%For a network with $s = 1$ or $r = 1$, due to monotonicity of objective and constraint, Algorithm \ref{algo} converges to the global optimum. When $s = 1$, for all $S_{\bt Q}$ we have $s_{\bt x}^* = P$, therefore,
%\begin{align*}
% \max_{ \substack{ S_{\bt Q}  \succeq \bt 0, \\ 0 \leq s_{\bt x} \leq P} }  \mathcal{L}( s_{\bt x},S_{\bt Q}, \mu) = \max_{ S_{\bt Q}}  \mathcal{L}( P, S_{\bt Q}, \mu).
%\end{align*}
%When $r = 1$, we have $s_{\bt Q}^*(S_{\bt X}) = \frac{s_{\bt Y_r \mid \bt Y_d}}{2^{C_0}-1}$ and
%\begin{align*}
%\max_{ \substack{ S_{\bt X} \succeq \bt 0, \ s_{\bt Q} > 0, \\ \mathrm{tr}(
%S_{\bt X}) \leq P} }   \mathcal{L}( S_{\bt X}, s_{\bt Q}, \mu) = 
% \max_{ \substack{S_{\bt X}, \\ \mathrm{tr}(S_{\bt X}) \leq P}} \mathcal{L}( S_{\bt X}, s_{\bt Q}^*, \mu).
%\end{align*}
%The obtained maximized rate is not jointly concave in $P$ and $C_0$ in general.

%For SISO Gaussian relay channel with an orthogonal additive Gaussian noise relay-destination link, joint optimization of transmission from source and relay is not a convex problem, and taking the convex envelope by time sharing can improve the achievable rate \cite{minenergy}. When the relay-destination link is digital, however, $R_{CF}\left(C_0\right)$ is a convex function and time sharing is not needed.

\subsection{Optimization of $S_{\bt X}$ for a Fixed $S_{\bt Q}$}
\label{section_opt_S_X}

Although the optimization problem (\ref{opt_prob}) is not concave 
in $S_{\bt X}$ for fixed $S_{\bt Q}$, we observe that the maximization of 
its Lagrangian (\ref{lagrangian}) at a given $\bar S_{\bt Q}$ is
concave for fixed $\mu \in (0, 1)$. Therefore, solving the maximization
\begin{equation}
\begin{array}{cl}
\underset{S_{ \bt X}}{\mathrm{maximize}} & \mathcal{L}( S_{\bt X}, \bar S_{\bt Q}, \mu) \\
\mathrm{subject~to} &  \mathrm{tr}( S_{\bt X}) \leq P,  \\
& S_{ \bt X} \succeq \bt 0,
\end{array}
\label{opt_S_X}
\end{equation}
using standard tools from convex optimization provides a global optimum of $S_{\bt Q}$ for the fixed $S_{\bt X}$; see the general sufficiency \cite[Proposition 3.3.4]{bertsekas}. To verify concavity, note that for fixed $S_{\bt Q}$ the Lagrangian can be written as a function of $S_{\bt X}$ as 
\begin{align}
& \mathcal{L}\left( S_{\bt X},\ \bar S_{\bt Q}, \ \mu \right) \nonumber \\ &= I(\bt X ; \widehat{ \bt Y}_R, \bt Y_D) - \mu I(\bt Y_R; \widehat{ \bt Y}_R \mid \bt Y_D) + \mu C_0  \nonumber \\ &=  \left(1-\mu \right) I(\bt X ; \widehat{ \bt Y}_R, \bt Y_D) + \mu
I(\bt X ; \bt Y_D) + \mathrm{const.} \nonumber \\ &= (1-\mu) \log \left|  H S_{\bt X}  H^{\dagger} + S_\bt{int} + \sigma^2 \bt I_{(r+d)} + 
\begin{bmatrix} \bar{S}_{\bt Q} &  \bt 0_{r\times d} \\
	\bt 0_{d \times r} & \bt 0_{d\times d} \end{bmatrix} \right| \nonumber \\ & \quad \quad + \mu \log \left| H_{SD} S_{ \bt X}  H_{SD}^{\dagger} +
S_\bt{int}^{\scriptscriptstyle (2,2)}+\sigma ^2\bt I_d \right| + {\rm const.},
\end{align}
which is a concave logdet function for $\mu \in (0,1)$.

The above form of the Lagrangian provides intuition about the
optimal choice of $S_{\bt X}$. The Lagrangian is a convex combination of two
terms. The first term corresponds to the channel from $\bt X$ to the combined
relay and destination receiver $(\widehat{ \bt Y}_R, \bt Y_D)$, while the
second term corresponds to the channel from $\bt X$ to the destination $\bt
Y_D$ alone. For larger values of $C_0$ (or equivalently small values of $\mu$), 
the optimal $S_{\bt X}$ should be close to the water-filling covariance matrix against the combined vector channel $H$. For small values of $C_0$, the optimal $S_{\bt X}$ should be close to the water-filling covariance matrix against the source-destination channel $H_{SD}$ alone. For a finite $C_0$, the optimal beamforming is obtained by maximizing the convex combination of the two terms.% returns the appropriate beamforming for the given $C_0$ through $\left(\bar{S}_{\bt Q}, \mu \right)$.

\subsection{Optimization of $S_{\bt Q}$ for a Fixed $S_{\bt X}$}
\label{section_opt_S_Q}

We now provide a closed-form solution for the $S_{\bt Q}$ that
maximizes the Lagrangian (\ref{opt_lagrangian}) for a given $\bar
S_{\bt X}$, i.e., the solution of 
\begin{equation}
\begin{array}{cl}
\underset{S_{ \bt Q}}{\mathrm{maximize}} & \mathcal{L}(\bar S_{\bt X},S_{\bt Q},\mu) \\
\mathrm{subject~to} &  S_{\bt Q}  \succeq \bt 0.
\end{array}
\label{opt_S_Q}
\end{equation}
%The key technique that makes a closed-form solution for global maximum possible is simultaneous diagonalization by *congruence \cite{horn}.
For the optimization of $S_{\bt Q}$ when $\bar S_{\bt X}$ is kept fixed, the
objective and constraint functions (\ref{opt_objective})-(\ref{opt_constraint}) 
can be rewritten as
\begin{align}
f_o &= \log \left| S_{ \bt Y_R \mid \bt Y_D} + S_{\bt Q} \right| - \log \left| S_{ \bt Y_R \mid \bt Y_D, \bt X} + S_{\bt Q} \right| + \mathrm{const.},
\label{opt_objective_Q}
\end{align}
\begin{align}
f_c = \log \left| S_{ \bt Y_R \mid \bt Y_D} + S_{\bt Q} \right| - \log \left| S_{\bt Q} \right| + \mathrm{const.},
\label{opt_constraint_Q}
\end{align}
and Lagrangian in (\ref{opt_S_Q}) can be rewritten as
\begin{multline}
\mathcal{L}( \bar S_{\bt X}, S_{\bt Q}, \mu)   = (1-\mu)  \log \left| S_{ \bt Y_R \mid \bt Y_D} + S_{\bt Q} \right| + \mu \log \left| S_{\bt Q} \right| \\ -  \log \left| S_{ \bt Y_R \mid \bt Y_D, \bt X} + S_{\bt Q} \right| + \mathrm{const.},
\label{lagrangian_for_Q}
\end{multline}
where the conditional covariances are obtained using the generalized Schur complement formula \cite{Rao}
%\begin{align}
\begin{multline}
S_{ \bt Y_R \mid \bt Y_D} \\ =  H_{SR}  S_{\bt X} H_{SR}^{\dagger} +
S_\bt {int}^{\scriptscriptstyle (1,1)} + \sigma^2 \bt I_r -  ( H_{SR}  S_{\bt X} H_{SD}^{\dagger} +
S_\bt {int}^{\scriptscriptstyle (1,2)} )  \\   ( H_{SD}  S_{\bt X} H_{SD}^{\dagger} + S_\bt {int}^{\scriptscriptstyle (2,2)}+ \sigma^2 \bt I_d  )^{-1}  ( H_{SD}  S_{\bt X} H_{SR}^{\dagger} + S_\bt {int}^{\scriptscriptstyle (2,1)} ), %\\ \nonumber
\end{multline}
%\end{align}
and
%\begin{multline}
\begin{align} 
%\begin{eqnarray*}
S_{ \bt Y_R \vert \bt Y_D, \bt X }   =&  H_{SR}  S_{\bt X} H_{SR}^{\dagger} + S_\bt {int}^{\scriptscriptstyle (1,1)} + \sigma^2 \bt I_r \nonumber \\ & -
  \begin{bmatrix}  H_{SR}  S_{\bt X} H_{SD}^{\dagger} + S_\bt {int}^{\scriptscriptstyle (1,2)} &  H_{SR}  S_{\bt X}  \end{bmatrix}  \nonumber \\ 
 & \cdot \begin{bmatrix} H_{SD}  S_{\bt X} H_{SD}^{\dagger} + S_\bt {int}^{\scriptscriptstyle (2,2)}+
\sigma^2 \bt I_d  &  H_{SD}  S_{\bt X}  \\  S_{\bt X} H_{SD}^{\dagger} & S_{\bt
X} \end{bmatrix}^{-1}  \nonumber \\ 
& \quad  \quad \cdot \begin{bmatrix} H_{SD}  S_{\bt X} H_{SR}^{\dagger} + S_\bt {int}^{\scriptscriptstyle (2,1)} \\  S_{\bt X} H_{SR}^{\dagger} \end{bmatrix}.
%\end{eqnarray*}
\end{align}
%\end{multline}
%Our goal is to maximize (\ref{lagrangian_for_Q}) over $S_{\bt Q}$.
%From (\ref{lagrangian_for_Q}) it can be seen that the optimal Lagrange multiplier is in the range of $\mu^* \leq 1$. This is true because for $\mu \geq 1$ the Lagrangian in (\ref{lagrangian_for_Q}) is non-positive, and the maximum is achieved by a $S_{\bt Q}$ with $\left| S_{\bt Q} \right| = \infty$. Therefore, penalizing the objective with $\mu > 1$ is the same as penalizing with $\mu = 1$.
%Note that the constraint $S_{\bt Q} \succeq 0$ is superfluous, since it is already implicit in the domain; if $|S_{\bt Q}|=0$, then $C_0=+\infty$.
The main step in obtaining the global optimum of (\ref{lagrangian_for_Q}) in closed form is the following simultaneous diagonalization by *congruence of $S_{ \bt Y_R \mid \bt Y_D, \bt X}$ and $S_{ \bt Y_R \mid \bt Y_D}$ based on \cite[Corollary 7.6.5]{horn}.

\begin {lemma} 
\label{lemma_diag}
There exists a non-singular matrix $ C_R \in \mathbb{C}^{r \times r}$
such that $ C_R^{\dagger} S_{ \bt Y_R \mid \bt Y_D, \bt X} C_R = \bt I_r$ 
and $C_R^{\dagger} S_{ \bt Y_R \mid \bt Y_D} C_R = \Lambda$, where
$\Lambda$ is a diagonal matrix. The diagonal elements $\lambda_i$'s are called the generalized eigenvalues, and we have $\lambda_{i} \geq 1$ for $i = 1, \dots, r$.  
\end{lemma}

\begin{IEEEproof} Both $S_{ \bt Y_R \mid \bt Y_D}$ and $S_{ \bt Y_R \mid \bt Y_D,
\bt X}$ are positive definite matrices. Let $S_{ \bt Y_R \mid \bt Y_D, \bt X}^{-1} = R^{\dagger}R$ be the Cholesky decomposition. Now, consider the eigendecomposition $RS_{ \bt Y_R \mid \bt Y_D}R^{\dagger} = V \Lambda V^{\dagger}$. Then $C_R = R^{\dagger}V$ satisfies $ C_R^{\dagger} S_{ \bt Y_R \mid \bt Y_D, \bt X} C_R = \bt I_r$ and $C_R^{\dagger} S_{ \bt Y_R \mid \bt Y_D} C_R = \Lambda$ simultaneously. Moreover, $S_{ \bt Y_R \mid \bt Y_D} \succeq  S_{\bt Y_R \mid \bt Y_D, \bt X}$ implies $\Lambda \succeq \bt I_r$.  \end{IEEEproof}

The above transformation makes elements of the relay's observed vector conditionally independent. We can now use the approach of \cite{cran,del} to solve the quantization noise optimization problem for describing the independent elements. For $\mu \in (0,1)$, the Lagrangian (\ref{lagrangian_for_Q}) can be 
written as
\begin{align}
\mathcal{L} \overset{(a)}{=}
& (1-\mu)\log \left| \Lambda \hat{S}_{\bt Q}^{-1} + \bt I_r \right| - \log
\left| \hat{S}_{\bt Q}^{-1} + \bt I_r \right| + \mathrm{const.} \nonumber \\ 
%%%%%%%%%%%%%%%%%%%%%%%%%%%%%%
\overset{(b)}{\leq}& (1-\mu)\log \left| \Lambda \Sigma_{\bt Q}^{-1} + \bt I_r \right| - \log \left| \Sigma_{\bt Q}^{-1} + \bt I_r \right| % \nonumber \\
	+ \mathrm{const.}
%%%%%%%%%%%%%%%%%%%%%%%%%%%%%%
\end{align}
where $(a)$ follows from the change of variable $\hat{S}_{\bt Q} = C_R^{\dagger}
S_{\bt Q} C_R$, with $C_R$ as in Lemma \ref{lemma_diag}, and $(b)$ follows from
\cite[Lemma 5]{cran} where $\Sigma_{\bt Q}$ comes from the eigendecomposition
$\hat{S}_{\bt Q} = U \Sigma_{\bt Q} U^{\dagger}$. Observe that the equality in
$(b)$ is obtained with $U = \bt I_r$.  Thus, it is without loss of optimality to
restrict $\hat{S}_{\bt Q}$ to be diagonal. 

Consider the change of variable \cite{del}
\begin{align} 
c_i = \log \left( 1 + \frac{\lambda_i}{\Sigma_{ \bt Q}^{ii}} \right), \ i = 1, \dots, r, \label{change_of_var}
\end{align}
where $\Sigma_{\bt Q}^{ii}$'s are the diagonal entries of $\Sigma_{\bt Q}$. An interpretation of $c_i$ is that it is the portion of the available $C_0$ allocated for quantization of the $i^{th}$ element of $C_R \bt Y_R$.
Using (\ref{change_of_var}), the Lagrangian can be written as 
\begin{align}
 \mathcal{L} = \sum_{i=1}^{r} \left(\left(1-\mu \right)c_i - \log \left( 2^{c_i} + \lambda_i- 1 \right)  \right) 
	+ \mathrm{const.}
\label{lagrangian_changed}
\end{align}
It can be readily checked that (\ref{lagrangian_changed}) is concave in $c_i$
for $\lambda_{i} \geq 1$. It is easy to see that the optimal $c_i^*$ is
\begin{align}
c_i^* = \left[\log(\lambda_i - 1) -  \log \frac{\mu}{1-\mu} \right]^+,
\label{c_i_opt}
\end{align}
therefore, $\Sigma_{\bt Q}^{ii,*}$ is given by
\begin{equation}
\Sigma_{\bt Q}^{ii,*} = \begin{cases} \frac{\mu}{1-
\frac{1}{\lambda_i} -\mu} & \mu < 1 - \frac{1}{\lambda_i} \\  +\infty
& \mu \geq 1 - \frac{1}{\lambda_i} \end{cases}.
\label{s_q_i_opt}
\end{equation}
The optimal $S_{\bt Q}$ is $S_{\bt Q}^* = C_R^{- \dagger}\Sigma_{\bt Q}^* C_R^{-1}$.

%\begin{figure}[t]
%\centering
%\input{myfig-3-v2.tex}
%\caption{Reverse water-filling for allocating $C_0$ among elements of $C_r \bt Y_r$ when $C_0 = \bar C_{0,5}$; the first component is asymptotically deterministic and the $6^{\mathrm{th}}$ component is reversely degraded. This happens, e.g., when $(s, d, r, t) = (5, 2, 6, 7)$.} \label{fig_rwf}
%\end{figure}

\begin{figure}[t]
\centering
\newcommand{\LineWidth}{1.5}
\newcommand{\fCSINR}{7}
\newcommand{\seCSINR}{2.1}
\newcommand{\tCSINR}{1.15}
\newcommand{\foCSINR}{-1.1}
\newcommand{\water}{-2.2}
\newcommand{\fifCSINR}{-2.2}
\newcommand{\siCSINR}{-5}

\begin{tikzpicture}

\begin{axis}[%
scale = 0.71,
width=\columnwidth,
height=0.9\columnwidth,
at={(0in,0in)},
scale only axis,
axis x line=middle,
xminorticks=false,
xmajorticks=false,
separate axis lines,
every inner x axis line/.append style={-, line width = \LineWidth pt},
every x tick label/.append style={rotate=-45,font=\color{black},line width=\LineWidth pt},
xmin=0,
xmax=12,
every axis x label/.style={ at={(ticklabel* cs:1.02)}, anchor=west},
xlabel=$i$,
%%%%%%%%%%%%%%%%%%%%%%%%%%%%%%%%%%%%%%%%%%%%%%%%%%%%%%%%%%%
axis y line=middle,
yminorticks=false,
ytick={ \fCSINR, \seCSINR, \tCSINR, \foCSINR, \water, \fifCSINR},
yticklabels={$\log\left(\lambda_1 - 1\right)$, $\log\left(\lambda_2 - 1\right)$, $\log\left(\lambda_3 - 1\right)$, $\log\left(\lambda_4 - 1\right)$, $\log {\frac{\mu}{1-\mu}}^{\color{white} \vdots \color{black}}$},
every inner y axis line/.append style={line width = \LineWidth pt},
every y tick label/.append style={font=\color{black},line width=\LineWidth pt, anchor=east},
ymin=-7,
ymax=10,
ylabel=$\log\left(\lambda_i - 1\right)$,
axis background/.style={fill=white}
]

\node[anchor=north] at (1,-0.2) {$1$};
\node[anchor=north] at (3,-0.2) {$2$};
\node[anchor=north] at (5,-0.2) {$3$};
\node[anchor=north] at (7,-0.2) {$4$};
\node[anchor=north] at (9,-0.2) {$5$};
\node[anchor=north] at (11,-0.2) {$6$};

\draw[line width=1pt]  (0,\fCSINR) -- (2,\fCSINR);
\draw[line width=1pt]  (2,\seCSINR) -- (4,\seCSINR);
\draw[line width=1pt]  (4,\tCSINR) -- (6,\tCSINR);
\draw[line width=1pt]  (6,\foCSINR) -- (8,\foCSINR);
\draw[line width=1pt]  (8,\fifCSINR) -- (10,\fifCSINR);
\draw[line width=1pt]  (10,\siCSINR) -- (12,\siCSINR);

\draw[line width=0.1pt]  (2,-0.3) -- (2,0.3);
\draw[line width=0.1pt]  (4,-0.3) -- (4,0.3);
\draw[line width=0.1pt]  (6,-0.3) -- (6,0.3);
\draw[line width=0.1pt]  (8,-0.3) -- (8,0.3);
\draw[line width=0.1pt]  (10,-0.3) -- (10,0.3);
\draw[line width=0.4pt]  (12,-0.3) -- (12,0.3);

\fill[pattern=north west lines] (0,\water) rectangle (2,\fCSINR);
\fill[pattern=north west lines] (2,\water) rectangle (4,\seCSINR);
\fill[pattern=north west lines] (4,\water) rectangle (6,\tCSINR);
\fill[pattern=north west lines] (6,\water) rectangle (8,\foCSINR);

\node[rotate = 90,anchor=east] at (11,-2.5) {$\dots$};
\node[anchor=north] at (11,\siCSINR-0.5) {$-\infty$};
\node[anchor=west] at (12,0) {$i$};

\end{axis}
\end{tikzpicture}%
\caption{Reverse water-filling for allocating $C_0$ among elements of $C_R \bt Y_R$ when $C_0 = \bar C_{0,5}$. The first component is asymptotically deterministic and the $6^{\mathrm{th}}$ component is reversely degraded. This happens, e.g., when $(s, d, r, t) = (5, 2, 6, 7)$.} \label{fig_rwf}
\end{figure}
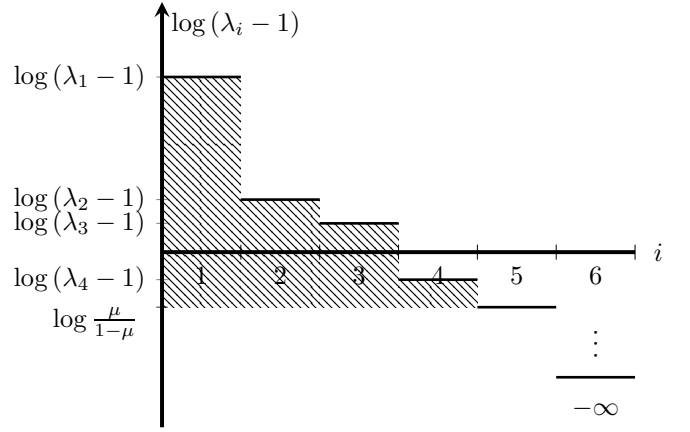

The above solution illustrates that the optimal compression involves independent quantization of elements of the relay's observed vector after transformation by matrix $C_R$ at quantization rates given by (\ref{c_i_opt}). This solution allocates quantization rates to elements of $C_R \bt Y_R$ using reverse water-filling on
$\log(\lambda_i - 1)$'s. An example of such a reverse water-filling
rate allocation is shown in Fig.~\ref{fig_rwf}. Here, $\lambda_i-1$
can be thought of as the $i^{\mathrm{th}}$ element's CSINR given 
$\bt Y_D$. To illustrate this interpretation of CSINR through an
example, consider the case of $r = 1$, where we have
\begin{align}
\lambda_1-1 = \frac{S_{\bt Y_R \mid \bt Y_D}-S_{ \bt Y_R \mid \bt Y_D, \bt X}}{S_{ \bt Y_R \mid \bt Y_D, \bt X}}.\end{align}
This is the ratio of the conditional variance of signal to the conditional variance of interference plus noise given $\bt Y_D$. In the next section, we explain how the magnitude of generalized eigenvalues determines the performance of MIMO compress-and-forward relaying.

A special case of this optimization problem is considered in
\cite{del}, where the noises at $\bt Y_R$ and $\bt Y_D$ are
independent and $S_{ \bt Y_R \mid \bt Y_D, \bt X}$ is an identity
matrix. In this case $S_{ \bt Y_R \mid \bt Y_D, \bt X}$ and $S_{ \bt
Y_R \mid \bt Y_D}$ can be diagonalized simultaneously by the CKLT of
$\bt Y_R$ given $\bt Y_D$, which is a unitary transformation
\cite{klt}. For the more general correlated noise case, the above
simultaneous diagonalization is needed to transform the matrix
optimization problem to scalar optimization.

We note that in \cite{JunChen} a diagonalization approach, known as CCA, is used to solve the same problem, but from a source coding perspective. Instead of diagonalizing $S_{ \bt Y_R \mid \bt Y_D, \bt X}$ and $S_{ \bt Y_R \mid \bt Y_D}$, the approach of  \cite{JunChen} diagonalizes $S_{\bt X \mid \bt Y_D}$ and $S_{\bt Y_R \mid \bt Y_D}$ using a singular value decomposition of the matrix $S_{\bt X \mid \bt Y_D}^{-1/2} K_{\bt X \bt Y_R} S_{\bt Y_R \mid \bt Y_D}^{1/2}$, where $K_{\bt X \bt Y_R}$ is a certain matrix of regression coefficients. It can be shown that the resulting diagonalization makes $S_{\bt X \bt Y_R \mid \bt Y_D}^{(1,2)}$ diagonal as well. Subsequently, the diagonal elements from the diagonalization of $S_{\bt X \bt Y_R \mid \bt Y_D}^{(1,2)}$ are used to find the optimal solution to the overall problem. 
We observe that the CCA approach in \cite{JunChen} can be interpreted as an indirect simultaneous diagonalization of $S_{\bt Y_R \mid \bt Y_D, \bt X}$ and $S_{ \bt Y_R \mid \bt Y_D}$, and is in fact equivalent to the transformation presented here. However, the direct simultaneous diagonalization of $S_{\bt Y_R \mid \bt Y_D, \bt X}$ and $S_{ \bt Y_R \mid \bt Y_D}$ is simpler and gives structural insight into the optimal MIMO compress-and-forward strategy. % in Section \ref{section_grevwf}.

\section{Interpreting the Generalized Eigenvalues}% in Rate Allocation}
\label{section_grevwf}

The optimization of quantization covariance in vector compress-and-forward involves reverse water-filling on $\log(\lambda_i - 1)$'s, where $\lambda_i - 1$ can be interpreted as the CSINR at the $i^{\mathrm{th}}$ element of $C_R \bt Y_R$. The reverse water-filling solution with respect to CSINRs reveals insight into the effectiveness of MIMO compress-and-forward relaying for improving the achievable rate.

%The numerical optimization of the compress-and-forward relaying strategy provides values of $\lambda_i - 1$, which can be interpreted as the CSINR at the $i^{\mathrm{th}}$ element of $C_R \bt Y_R$. The reverse water-filling solution with respect to CSINRs reveals insight into the effectiveness of MIMO compress-and-forward relaying for improving the overall rate.

%The reverse water-filling solution with respect to the generalized eigenvalues of the conditional covariance matrices reveals significant insight on the effectiveness of MIMO compress-and-forward relaying for improving the overall rate, as explained below. 

\subsection{The Slope of $\bar R_{CF}(C_0)$}
\label{section_slope}

For the MIMO relay channel under study, the following result relates
the slope of $\bar R_{CF}(C_0)$ to the magnitude of the generalized
eigenvalues $\lambda_i$'s. 

\begin{lemma}
In the optimal allocation of the relaying link capacity $C_0$ to elements of $C_R \bt Y_R$ for quantization, given in (\ref{c_i_opt}), let $\bar C_{0,i}$ be the largest $C_0$ for which the optimal quantization rate of the $i^\mathrm{th}$ element is zero, i.e., $c_i^* > 0$ if and only if $C_0 > \bar C_{0,i}$. We have 
\begin{align}
\bar C_{0,i} = \sum_{j = 1}^{i-1} \log \frac{\lambda_j - 1}{\lambda_i - 1}, \ \ i = 2, \dots, r,
\label{crit_C_0}
\end{align}
and
\begin{align}
\left.\frac{d \bar R_{CF}(C_0)}{d C_0}\right|_{C_0 = \bar C_{0,i}} = 1 - \frac{1}{\lambda_i},
\label{crit_slope}
\end{align}
where $\bar R_{CF}(C_0)$ is (\ref{opt_prob}) evaluated at given $\bar S_{\bt X}$.
\label{theorem_slope}
\end{lemma}

\begin{IEEEproof} 
By the result of Section \ref{section_opt_S_Q}, the optimization problem
(\ref{opt_S_Q}) can be transformed into a convex problem, therefore strong
duality holds. Moreover, the optimized $\bar R_{CF}(C_0)$ is differentiable over $C_0 > 0$. The slope of $\bar R_{CF}(C_0)$ at $C_0$ is in fact the optimal Lagrange multiplier in 
the KKT solution at the given $C_0$ \cite{boyd}, i.e., 
\begin{align}
\frac{d \bar R_{CF}(C_0)}{dC_0} = \mu^*(C_0).
\label{slope}
\end{align}
%For $\bar C_{0,i-1} \leq C_0 \leq \bar C_{0,i}$, $i = 2, \dots, r$, i.e., 
When $C_0$ is such that $c_{i}^* = 0$ and $c_1^* + \dots + c_{i-1}^* =
C_0$ for $2 \leq i \leq r$, using the KKT solution (\ref{c_i_opt}),
the Lagrange multiplier is given by 
\begin{align}
\mu^*(C_0) = \frac{1}{1+\left(\frac{2^{C_0}}{\prod_{j =
1}^{i-1}(\lambda_j - 1)}\right)^{\frac{1}{i-1}}}.
\label{optlag}
\end{align}
Now, $\bar C_{0,i}$ is the largest value of $C_0$ for which the optimal quantization rate of the $i^\mathrm{th}$ element is zero, $c_{i}^* = 0$. 
Using (\ref{c_i_opt}) again, we have 
\begin{align}
\log \frac{1-\mu^*(\bar C_{0,i})}{\mu^*(\bar C_{0,i})} + \log(\lambda_i - 1) = 0.
\end{align}
This together with (\ref{slope}) and (\ref{optlag}) yields (\ref{crit_C_0}) and (\ref{crit_slope}).
\end{IEEEproof}

%\begin{figure}[t]
%\centering
%\input{myfig-2.tex}
%\caption{The optimized compress-and-forward rate for fixed $S_{\bt X}$ as a function of $C_0$; the slope of the curve at $\bar C_{0,i}$ is $1-\frac{1}{\lambda_i}$.} \label{fig_slope}
%\end{figure}

\begin{figure}[t]
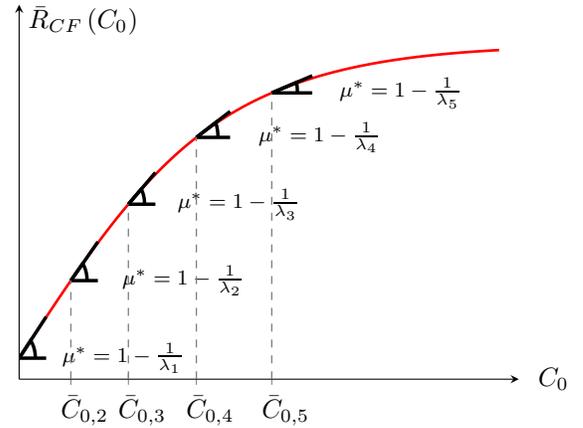

\centering
\newcommand{\LineWidth}{2}

%
\caption{For a fixed $S_{\bt X}$, the slope of the optimized compress-and-forward rate curve at $\bar C_{0,i}$ is $1-\frac{1}{\lambda_i}$ for $i = 1, \cdots, r$.}
\label{figslope}
\end{figure}

Fig.~\ref{figslope} illustrates this result. The slope of the
compress-and-forward achievable rate at various points of $C_0$ is
determined by the generalized eigenvalues of the conditional
covariance matrices. A larger $\lambda_i$ implies a slope closer to
the maximum value of 1.

% for $(s, d, r, t) = (5, 2, 6, 7)$ case.

\subsection{Zero CSINR: Reversely Degraded Components}
\label{section_revdeg}

When $\lambda_i(S_{\bt Y_R \mid \bt Y_D}S_{ \bt Y_R \mid \bt Y_D, \bt X}^{-1}) = 1$, the corresponding element in $C_R \bt Y_R$ is conditionally independent of $\bt X$ given $\bt Y_D$. Since such an element does not convey any further information to destination, the optimized compress-and-forward does not assign any portion of $C_0$ for describing it. %We call this the reversely degraded component of the MIMO relay channel.%A general relay channel for which the Markov chain $\bt X \leftrightarrow \bt Y_d \leftrightarrow \bt Y_r $ holds, i.e., $\bt Y_r$ given $\bt Y_d$ is equal in distribution to $\bt Y_r$ given $(\bt Y_d, \bt X)$, is called a reversely degraded channel. In such a channel the relay cannot help and the capacity is the capacity of the direct transmission \cite[Theorem 2]{caprelay}. This motivates the following definition.

\begin{definition}
The $i^{\rm th}$ element of $C_R \bt Y_R$ is called reversely degraded, if $\lambda_i(S_{\bt Y_R \mid \bt Y_D}S_{ \bt Y_R \mid \bt Y_D, \bt X}^{-1}) = 1$.
\label{rev_deg_def}
\end{definition}

When the relay node is equipped with a large number of antennas, the
number of elements of $C_R \bt Y_R$ to be quantized, i.e., the elements
that are not reversely degraded, can be much smaller than the number of relay antennas. The following proposition bounds the number of reversely
degraded components in the relay channel. It shows that the optimal
scheme quantizes no more than $\min(r, s^\mathrm{r})$ elements. 
This result is used for the constant gap characterization of
the capacity in Section \ref{section_cg} as well.  

\begin{theorem}
\label{prop_rev_deg}
In the MIMO relay channel under study, the number of reversely
degraded components is at least $(r-s^{\mathrm{r}})^+$, where
$s^{\mathrm{r}}$ is the number of independent data streams transmitted
by the source, i.e., $\mathrm{rank}(\bar S_{\bt X}) = s^\mathrm{r}$.
Hence, the optimal quantization scheme in (\ref{c_i_opt}) describes at
most $\min (r, s^{\mathrm{r}})$ elements of $C_R \bt Y_R$.
\end{theorem}

\begin{IEEEproof}
By Lemma \ref{lemma_diag}, $C_R \left(S_{\bt Y_R \mid {\bt
Y}_D} - S_{{\bt Y}_R \mid {\bt Y}_D, \bt X}\right) C_R^\dagger =
\Lambda - \bt I_r$. Therefore, the number of reversely degraded
components of the channel is equal to $r - \mathrm{rank}(C_R
\left(S_{\bt Y_R \mid {\bt Y}_D} - S_{{\bt Y}_R \mid {\bt Y}_D, \bt
X}\right) C_R^\dagger) = r - \mathrm{rank}(S_{\bt Y_R \mid {\bt Y}_D}
- S_{{\bt Y}_R \mid {\bt Y}_D, \bt X})$. Now, using the generalized
Schur complement formula, we have 
\begin{align}
S_{\bt Y_R \mid {\bt Y}_D} - S_{{\bt Y}_R \mid {\bt Y}_D, \bt X} = S_{{\bt Y}_R, \bt X \mid {\bt Y}_D}^{(1,2)} S_{\bt X \mid {\bt Y}_D}^{-1} S_{{\bt Y}_R, \bt X \mid {\bt Y}_D}^{(2,1)}.
\end{align}
The result follows by noting that $\mathrm{rank}(S_{\bt Y_R \mid {\bt Y}_D} - S_{{\bt Y}_R \mid {\bt Y}_D, \bt X}) \leq \min(r , s^\mathrm{r})$.
\end{IEEEproof}

\subsection{Infinite CSINR: Asymptotically Deterministic Components}
\label{section_detcomp}

Relaying can be particularly effective if the relay observation $\bt Y_R$ is a
deterministic function of $(\bt Y_D, \bt X)$. For example, this can happen asymptotically in a SISO relay channel with $(s, d, r, t) = (1, 1, 1, 1)$, where we have
\begin{align}
\bt Y_R = \left(h_{SR} - \frac{h_{TR}h_{SD}}{h_{TD}}\right) \bt X + \frac{h_{TR}}{h_{TD}} \bt Y_D + \bt N_1 - \frac{h_{TR}}{h_{TD}} \bt N_2,
\end{align}
assuming $h_{TD} \neq 0$ and $h_{SR}h_{TD} \neq h_{SD} h_{TR}$. When the noise power is zero, we have $\bt Y_R = f(\bt X, \bt Y_D)$. In a relay channel with this type of determinism, compress-and-forward achieves the cut-set upper bound \cite{coverkim}, and every relay bit improves the overall transmission rate by exactly one bit (if the relaying link capacity is not too large). In the above example, as $\sigma^2 \to 0$, the slope of $\bar R_{CF}(C_0)$ at small $C_0$ asymptotically approaches 1; this can be verified by noting that
\begin{align}
\lim_{\sigma^2 \to 0} S_{ \bt Y_R | \bt Y_D} = \frac{\left|h_{SR} h_{TD}-{h_{SD}h_{TR}}\right|^2 P}{\left|h_{SD}\right|^2+\left|h_{TD}\right|^2}.
\end{align}
and
\begin{align}
\lim_{\sigma^2 \to 0} S_{ \bt Y_R | \bt Y_D, \bt X} = 0.
\end{align}
Thus, $\lim_{\sigma^2 \to 0} \lambda_1 = \infty$. Hence, $\lim_{\sigma^2 \to 0} \left.\frac{d \bar R_{CF}(C_0)}{d C_0}\right|_{C_0 = 0} = 1$ by Lemma \ref{theorem_slope}. The following result reveals the number of asymptotic deterministic
components of this type for the MIMO relay channel under consideration. 
%In the MIMO relay channel, when the $i^\mathrm{th}$ element is asymptotically deterministic, the slope of $\bar R_{CF}$ at $\bar C_{0,i}$ given in (\ref{crit_slope}) approaches 1 as $\sigma^2 \to 0$.  Next proposition counts the number of asymptotically deterministic components of the MIMO relay channel. 

%Under certain conditions on the numbers of antennas, a generalized eigenvalue $\lambda_i$ goes to infinity as the power of background noise $\sigma^2$ goes to zero. 
%In Theorem \ref{theorem_slope}, we learned that the slope of $\bar R_{CF}(C_0)$ curve is increasing in the generalized eigenvalues. This motivates the following definition.

\begin{definition}
The $i^{\mathrm th}$ element of $C_R \bt Y_R$ is called asymptotically
deterministic, if $\lim_{\sigma^2 \to 0} \lambda_i(S_{\bt Y_R \mid \bt
Y_D}S_{ \bt Y_R \mid \bt Y_D, \bt X}^{-1}) = \infty$.
\label{det_def}
\end{definition}

%In general, when observation of the relay approaches a deterministic function of observation of destination and source signal, i.e., when $\bt Y_r \to f(\bt Y_d, \bt X)$ as $\sigma^2 \to 0$, for small $C_0$ increasing $C_0$ increases the overall throughput in a 1:1 ratio asymptotically.

%In \cite{coverkim} it has been shown that in a relay channel whenever $\bt Y_r$ is a deterministic function of $(\bt Y_d, \bt X)$, compress-and-forward achieves the cut-set bound, and for $C_0$ smaller than $I( \bt X ; \bt Y_r , \bt Y_d ) - I( \bt X ; \bt Y_d )$, evaluated at $\arg \max_{S_{\bt X}} I( \bt X ; \bt Y_d )$, increasing $C_0$ increases the overall throughput in a 1:1 ratio. This motivates the following definition.

%\begin{lemma}
%We have $\lim_{\sigma^2 \to 0} \lambda_i(S_{\bt Y_r \mid \bt Y_d}S_{ \bt Y_r \mid \bt Y_d, \bt X}^{-1}) = \infty$ if and only if $s^\mathrm{r} > i - 1$, $r + d > t + i - 1$, and $d < s^\mathrm{r} + t - (i - 1)$.
%\label{lemma_det}
%\end{lemma}

\begin{theorem}
In the MIMO relay channel under study, the number of asymptotically
deterministic components is
\begin{equation}
r' - r'' = \min \left( r, s^\mathrm{r}, (r+d-t)^+, (s^\mathrm{r}+t-d)^+ \right),
\label{det_comps}
\end{equation}
almost surely, where 
\begin{align}
	r' &\triangleq \mathrm{rank}(S_{\bar{\bt Y}_R \mid \bar{\bt Y}_D}) = \min \left( r , (s^\mathrm{r}+t-d)^+ \right), \label{r'} \\
	r'' &\triangleq \mathrm{rank}(S_{\bar{\bt Y}_R \mid \bar{\bt Y}_D, \bt X})= \min \left( r , (t-d)^+ \right) \label{r''}.
\end{align}
Here,
\begin{align}
	\bar{\bt Y}_R &\triangleq H_{SR}\bt X + H_{TR}\bt X_t \label{r_nonoise}, \\
	\bar{\bt Y}_D &\triangleq H_{SD}\bt X +H_{TD}\bt X_t  \label{d_nonoise},
\end{align}
are the noiseless parts of the relay and destination's observations, respectively. 
Hence, for $i = 1, \dots, r'-r''$, the slope (\ref{crit_slope}) approaches 1 as $\sigma^2 \to 0$. %Moreover,
%\begin{align}
%r' =	\mathrm{rank}(S_{\bar{\bt Y}_r \mid \bar{\bt Y}_d}) &= \min \left( r , (s^\mathrm{r}+t-d)^+ \right) \label{r_nonoise} \\
%r'' = \mathrm{rank}(S_{\bar{\bt Y}_r \mid \bar{\bt Y}_d, \bt X}) &= \min \left( r , (t-d)^+ \right) \label{d_nonoise}
%\end{align}
\label{thm_det_comp}
\end{theorem}

\begin{IEEEproof} See Appendix \ref{app_prf_thm_det_comp}.
\end{IEEEproof}

A key implication of the above result is the following. For the MIMO
relay channel under consideration, relaying with small $C_0$ is most effective if $\left.\frac{d \bar R_{CF}(C_0)}{d C_0}
\right|_{C_0 = 0}$ is at the maximum value of 1.
This happens when $\lambda_1 \to \infty$, i.e., when the MIMO relay channel has
at least one asymptotic deterministic component, or equivalently, by Theorem 
\ref{thm_det_comp}, when $r' > r''$. By (\ref{det_comps}), this holds when $r + d
> t$ and $d < s^{\rm r}+t$. 
%Intuitively, $\left.\frac{d \bar R_{CF}(C_0)}{d > C_0}\right|_{C_0 = 0} = 1$, if and only if, first, $d < s^{\rm r}+t$, i.e., the number of destination antennas alone cannot completely remove interference without the help of the relay, and second, $r+d > t$, i.e., combining relay and destination antennas is able to completely mitigate the total number of interference dimensions.
We state this result below and will return to the interpretation of
this condition in Section \ref{section_connection}.

\begin{corollary}
For the MIMO relay channel under study, 
$\lim_{\sigma^2 \to 0} \left.\frac{d \bar R_{CF}(C_0)}{d C_0}\right|_{C_0 = 0} = 1$ if and only if $r+d>t$ and $d<s^{\rm r}+t$.
\end{corollary}

\section{DoF Improvement by Relaying}
\label{section_dof}

We gain further insight into the benefit of relaying by analyzing DoF of the MIMO relay channel. Consider the DoF of the channel under study without the relay, and with a relay that has infinite link capacity, i.e., respectively, 
\begin{align}
DoF_D &\triangleq \lim_{\rho \to \infty} \frac{R_{CF}(0)}{\log \rho} =
\min\left( s, (d-t)^+ \right),  \label{dof_d} \\
DoF_R &\triangleq \lim_{\rho \to \infty} \frac{R_{CF}(\infty)}{\log \rho} = \min\left( s, (r+d-t)^+ \right). \label{dof_r} 
\end{align}
At infinite relay link capacity, the DoF gain due to relaying is
\begin{equation}
DoF_R - DoF_D = \min \left( r, s, (r+d-t)^+, (s+t-d)^+ \right).
\label{dof_diff}
\end{equation}
Curiously, the above expression is identical to the number of
asymptotically deterministic components derived in the previous
section. We will return to this connection in the next section.
%The above result characterizes the DoF improvement with a relay 
%link of infinite capacity. The DoF characterization below considers

We now characterize the overall DoF gain due to relaying. First, we note that the design of the combining matrix at the relay is crucial for achieving the optimal DoF. Without any combining, naive i.i.d. quantization of the relay's observed vector on a per-antenna basis results in a DoF loss in general.

\begin{theorem}
\label{thm_sub_opt_dof}
For the relay channel under study, the achievable DoF improvement by compress-and-forward with relay's quantization scheme (\ref{testchannel}) restricted to $\bt Q \sim \mathcal{CN} (\bt 0_{r \times 1}, q \bt I_r)$ is
%\begin{align}
%\Delta DoF_{i.i.d.} =& \min \left( 1, \frac{\alpha}{r'} \right) \nonumber \\ &\left( \min\left( s, (r+d-t)^+ \right) - \min\left( s , (d-t)^+ \right) \right)
%\label{dof_iid}
%\end{align}
\begin{align}
\Delta DoF_{i.i.d.} = \left( DoF_R - DoF_D \right) 
	\min \left( 1, \frac{\alpha}{r'} \right)
\label{dof_iid}
\end{align}
almost surely, where
$r' = \mathrm{rank}(S_{\bar{\bt Y}_R \mid \bar{\bt Y}_D}) = \min(r, (s+t -d)^+)$ and $\alpha$ is the DoF of the relaying link capacity $C_0$ as in (\ref{c0}).
%\begin{align}
%DoF_R &\triangleq \min\left( s, (r+d-t)^+ \right), \label{dof_r} \\
%DoF_D &\triangleq \min\left( s, (d-t)^+ \right)  \label{dof_d}.
%\end{align}
\end{theorem}

\begin{IEEEproof} See Appendix \ref{app_prf_thm_sub_opt_dof}.
\end{IEEEproof}

%The optimized compress-and-forward achieves cut-set bound to within a constant additive gap. Therefore, it increases the overall DoF in a 1:1 ratio of relay-destination's link DoF, $\alpha$. 

Note that when $d<t$, the coefficient of $\alpha$ in (\ref{dof_iid}) is 
\begin{align}
\frac{(DoF_R - DoF_D)}{r'} = 
\frac{\min \left( s, r + d - t \right)}{\min \left( r, s+t-d \right)} < 1,
\end{align}
whereas, by the next theorem, the coefficient of $\alpha$ in the optimal DoF gain is 1. This loss in DoF is due to the non-optimized choice of quantization scheme.

The optimal DoF can be achieved by using a combining matrix $\widetilde{C}_R$ of dimension $\min(r,s,(r+d-t)^+) \times r$ followed by i.i.d. quantization at the relay. The combining matrix aligns the observed interference at the relay with the row space of the observed interference at the destination. The optimality of this scheme follows by the cut-set bound (\ref{csb_max_min}). Details of the proof are deferred to Appendix \ref{app_prf_thm_opt_dof}. We call this combining strategy distributed zero-forcing, because in zero-forcing of interference with $r + d$ antennas pooled together, the combining matrix $\widetilde{C}_R$ is the submatrix corresponding to $r$ antennas of the relay.

\begin{theorem} 
For the relay channel under study, the optimal DoF improvement by relaying is
\begin{align}
\Delta DoF^* = \min \left( DoF_R - DoF_D, \alpha \right)
\label{opt_dof}
\end{align}
almost surely.
\label{thm_opt_dof}
\end{theorem}

\begin{IEEEproof} See Appendix \ref{app_prf_thm_opt_dof}.
\end{IEEEproof}

%We defer the details of the proof to appendix, but note that by cut-set bound (\ref{csb_max_min}), the above DoF gain is the maximum possible with relaying. To show that this DoF gain is  achievable, instead of using the optimized relay quantization matrix, it is possible to design a simpler combining matrix $\widetilde{C}_r$ of dimension $\min(r,s,(r+d-t)^+) \times r$ followed by i.i.d. quantization at the relay to based on the principle of aligning the observed interference at the relay with the row space of the observed interference at the destination. We call such a relaying strategy distributed zero-forcing, because it essentially concentrates all the interference into a subspace while making the rest of the subspace interference free. 

%Therefore, even from DoF perspective, i.i.d. quantization is not optimal in general.%

\section{Connection Between DoF Improvement and Asymptotically Deterministic Components}

\label{section_connection}

Combining the results of the previous two sections, we see that the DoF improvement at infinite relay link rate
is equal to the number of asymptotically deterministic
components in the MIMO relay channel. Thus, the DoF
improvement at $C_0 = \infty$ is related to the slope of the
achievable rate at small $C_0$. In this section, we first provide an
information theoretical proof of this equivalence, then further 
explore this connection via the DoF-optimal distributed zero-forcing
combiner design at the relay.

\begin{theorem}
For the MIMO relay channel under study, the maximum DoF improvement
achieved with an infinite capacity relay link is equal to the number
of asymptotically deterministic components, i.e., 
\begin{equation}
DoF_R - DoF_D = r'-r''.
\label{chain_dof}
\end{equation}
The DoF gain, or equivalently the number of asymptotically
deterministic components, is greater than zero if and only if $d < s + t $ and $r + d > t$. 
\label{thm_finite_infinite}
\end{theorem}
\begin{IEEEproof} Fix $S_{\bt X}$ to be full-rank. By Theorem
\ref{thm_det_comp}, the number of asymptotically deterministic
components is $r'-r''$, the difference of ranks of $S_{\bar{\bt Y}_R \mid \bar{\bt Y}_D}$ and 
$S_{\bar{\bt Y}_R \mid \bar{\bt Y}_D, \bt X}$.
To relate this to $DoF_R - DoF_D$, we expand the conditional mutual information below in two different ways
\begin{align}
I(\bt X ; \bt Y_R \mid \bt Y_D) = I(\bt X ; \bt Y_R, \bt Y_D) - I(\bt X ; \bt Y_D),
\end{align}
\begin{align}
I(\bt X ; \bt Y_R \mid \bt Y_D) = h(\bt Y_R | \bt Y_D) - h(\bt Y_R | \bt Y_D, \bt X).
\end{align}
The DoF of the first equality is $DoF_R - DoF_D$, while the DoF of the second is $r'-r''$.

%Similar to above, DoF of the left hand side is $r'-r''$. For the right hand side, let $r_d \triangleq \mathrm{rank}(S_{\bar{\bt Y}_d}) = \min (d, s+t)$ and $r_{d \mid s} \triangleq \mathrm{rank}(S_{\bar{\bt Y}_d \mid \bt X})= \min (d, t)$ to have
%\begin{align*}
%I(\bt X ; \bt Y_d) &=\log \frac{\left|S_{\bt Y_d}\right|}{\left|S_{\bt Y_d \mid \bt X}\right|} \\ & = \sum_{i = 1}^{r_{d \mid s}} \log \frac{\lambda_{i}(S_{\bar{\bt Y}_d})+\sigma^2}{\lambda_{i}(S_{\bar{\bt Y}_d \mid \bt X})+\sigma^2} + \\ & \sum_{i = r_{d \mid s}+1}^{r_d} \log \frac{\lambda_{i}(S_{\bar{\bt Y}_d})+\sigma^2}{\sigma^2}, 
%\end{align*}
%which has DoF equal to $r_d - r_{d \mid s} = \min (s, (d-t)^+)$. Similarly, DoF of $I(\bt X ; \bt Y_r, \bt Y_d)$ term is given by $\min (s, (r+d-t)^+)$. This completes the proof.
\end{IEEEproof}

%\label{section_ddof}

%In section \ref{section_opt_S_Q}, we learned that the optimal quantization scheme for vector compress-and-forward is to combine $\bt Y_r$ using the simultaneous diagonalization matrix then describe $\min(r, s^\mathrm{r})$ best elements of $C_r \bt Y_r$ with independent quantization noise. In this section, we show that the DoF-optimal combining matrix involves distributed zero-forcing of interference. Moreover, describing $\min(r, s, \left(r+d-t\right)^+)$ elements with i.i.d. quantization noises is DoF-optimal. %The DoF analysis in this section provides conditions on $r$ and $\alpha$ for improving DoF by interference mitigation and signal enhancement using a relay.

%Before discussing the DoF-optimal scheme, we first show that describing all $r$ elements of $\bt Y_r$ with i.i.d. quantization noises wastes the DoF of relay-destination link $\alpha$.  A simple quantization scheme is to describe all $r$ elements of $\bt Y_r$ with independent quantization noise.

%saving alpha by combining 
%r and alpha needed for DoF gain
%dimension reduction for DoF gain

We now further explore this connection via the concept of
determinism in the relay channel. As already mentioned, for a relay channel with $\bt Y_R = f\left(\bt Y_D, \bt X\right)$, compress-and-forward improves the rate in a 1:1 ratio of $C_0$ as long
as $I\left(\bt X ; \bt Y_R, \bt Y_D \right) > I\left(\bt X ; \bt Y_D
\right)$ for some $p(\bt x)$ \cite{coverkim}. A relay channel with
$\bt Y_R = f\left(\bt Y_D\right)$, however, is reversely degraded and
its capacity, given by $\max_{p\left(\bt x\right)} I\left(\bt X ; \bt
Y_D \right)$, does not depend on $C_0$ \cite{caprelay}.

%In the MIMO relay channel considered in this paper the deterministic components that are not reversely degraded

%The components of $Y_r$ that is $Y_r = f\left(Y_d, X\right)$ but not $Y_r = f\left(Y_d\right)$ can be revealed

For the MIMO relay channel considered here, when $d \geq t + s$, 
the channel (\ref{r_nonoise})-(\ref{d_nonoise}) is reversely degraded,
because 
\begin{align}
\bar{\bt Y}_R = \begin{bmatrix}H_{SR}  H_{TR}\end{bmatrix} \left(\begin{bmatrix}H_{SD}^\dagger \\ H_{TD}^\dagger\end{bmatrix} \begin{bmatrix}H_{SD}  H_{TD}\end{bmatrix}\right)^{-1}\begin{bmatrix}H_{SD}^\dagger \\ H_{TD}^\dagger\end{bmatrix}\bar{\bt Y}_D.
\label{det_b}
\end{align}
In this case, the slope of $\bar R_{CF}(C_0)$ does not approach one.

When $d < t + s$, the channel (\ref{r_nonoise})-(\ref{d_nonoise}) can contain deterministic components that are not reversely degraded. It turns out that the DoF-optimal distributed zero-forcing scheme mentioned in Section \ref{section_dof} can reveal such deterministic components. More specifically, we first form $\bar{\bt Y}'_R$, defined to be the component of $\bar{\bt Y}_R$ that is independent of $\bar{\bt Y}_D$, i.e.,
\begin{multline}
\bar{\bt Y}'_R = \bar{\bt Y}_R - S_{\bar{\bt Y}_R, \bar{\bt Y}_D}^{(1,2)} S_{\bar{\bt Y}_D}^{-1} \bar{\bt Y}_D  \\ = \begin{bmatrix}H_{SR}  H_{TR}\end{bmatrix} \bar{P}\begin{bmatrix}\bt X \\ \bt X_t\end{bmatrix} \triangleq \begin{bmatrix}\bar{H}_{SR}  \bar{H}_{TR}\end{bmatrix}\begin{bmatrix}\bt X \\ \bt X_T\end{bmatrix},
\end{multline}
where
\begin{align}
\bar{P} = \bt I_{s+t} - \begin{bmatrix}S_{\bt X} H_{SD}^\dagger \\ S_{\bt X_T} H_{TD}^\dagger\end{bmatrix}S_{\bar{\bt Y}_D}^{-1} \begin{bmatrix}H_{SD}  H_{TD}\end{bmatrix}.
\end{align}
is a projection matrix. Now, consider the effective channel
\begin{align}
\begin{bmatrix}\bar{\bt Y}'_R \\ \bar{\bt Y}_D\end{bmatrix} = \begin{bmatrix}\bar{H}_{SR}  \bar{H}_{TR} \\ H_{SD}  H_{TD} \end{bmatrix}\begin{bmatrix}\bt X \\ \bt X_T\end{bmatrix}.
\end{align}
If $r+d > t$, then we can select a combining matrix $\bar{C}_R \in \mathbb{C}^{(r+d-t)^+ \times r}$ such that
\begin{align}
\begin{bmatrix}\bar{C}_R \ \bar{A} \end{bmatrix}\begin{bmatrix}\bar{H}_{TR} \\ H_{TD} \end{bmatrix}= \bt 0,
\end{align}
in which case
\begin{align}
\bar{C}_R\bar{\bt Y}'_R = \begin{bmatrix}\bar{C}_R \ \bar{A} \end{bmatrix}
\begin{bmatrix}\bar{H}_{SR}  \\ H_{SD} \end{bmatrix}\bt X - \bar{A} \bar{\bt Y}_D
\end{align}
is the component of $\bar{\bt Y}_R$ that is a function of $(\bar{\bt
Y}_D, \bt X)$, but not a function of $\bar{\bt Y}_D$ alone. 
Such deterministic component gives rise to the unit slope of
$\bar{R}_{CF}(C_0)$ at $C_0=0$. It exists if and only if both $r+d > t$ and $d < s+t$.

We now see that using distributed zero-forcing combining at the relay not only gives rise to DoF-optimal relaying strategy, but also reveals the deterministic components of the MIMO relay channel. It is for this reason that the condition for having DoF
improvement by $C_0 = \infty$ is the same as the condition under
which relay has deterministic components and has
$\lim_{\sigma^2 \to 0} \frac{dR_{CF}(C_0)}{dC_0} = 1$ at small $C_0$. 

From the DoF point of view, one can interpret the above conditions on
the numbers of antennas in the following way. To improve DoF by adding
$r$ more antennas to the destination, the destination must not already
have sufficient antennas to be able to completely mitigate interference
and to resolve the intended signal, hence $d < s + t$ is required.
Further, the relay and the destination together must have enough
antennas to mitigate all of the interference, so $r + d > t$ is
needed. Thus, we need $r+d>t$ and $d<s+t$, as otherwise adding $r$
more antennas to the receiver cannot improve the spatial DoF.

%Proposition \ref{thm_det_comp} shows that the number of asymptotically deterministic components is $r'-r''$. The DoF improvement at $C_0 = \infty$ is $DoF_R - DoF_D$. Equality of these two in (\ref{chain_dof}) follows by the chain rule for mutual information.% $r' -r'' = \min \left(s , \left(r+d - t\right)^+ \right) - \min \left(s , \left(d - t\right)^+ \right)$ can be seen from the chain rule
%\begin{align*}
%I(\bt X ; \bt Y_r \mid \bt Y_d) = I(\bt X ; \bt Y_r, \bt Y_d) - I(\bt X ; \bt Y_d),
%\end{align*}
%by calculating DoF of both sides of equality.

%Before discussing the DoF-optimal scheme, we first characterize the achievable DoF by describing all $r$ elements of $\bt Y_r$ with i.i.d. quantization. This simple quantization scheme might be of practical interest. However, we observe that when $d < t$ it is wasteful of relay-destination link's DoF, $\alpha$.

\section{Constant Gap Characterization of Capacity}
\label{section_cg}

Throughout this paper, we have focused on the compress-and-forward
relaying strategy with Gaussian input and quantization. In this final section of the paper, we show that this relaying strategy achieves the capacity of the Gaussian MIMO relay channel with noise correlation to within an additive gap that only depends on the number of antennas. 

%The following theorem shows that for the Gaussian MIMO relay channel with noise correlation a suboptimal evaluation of compress-and-forward in (\ref{cf2}) is to within $\min(r, s)$ b/sec/Hz of the capacity. As a corollary, we argue that the KKT solution obtained by algorithm \ref{algo} achieves the capacity of this channel to within the same additive gap for all values of channel parameters.

\begin{theorem}
For the MIMO relay channel under study, the achievable rate
using optimized input and quantization covariance matrices given by
Algorithm \ref{algo} is to within $\min(r,s)$ bits of the capacity.
\label{thm_cg}
\end{theorem}

\begin{IEEEproof} See Appendix \ref{app_prf_thm_cg}.
\end{IEEEproof}

\section{Numerical Examples}
\label{section_sim}

In this section, we provide numerical examples of MIMO relaying in a wireless cellular environment. We simulate a picocell network with a BS transmitting at a maximum power of 1W over 10MHz to a user located $100$m away. A relay node, located $10$m away from the user, pools antennas with user's receiver over a digital link of capacity $C_0$. Here, $C_0$ is normalized by the 10MHz of bandwidth and is expressed in b/s/Hz. Interfering BSs are place on a hexagonal grid, with minimum BS-to-BS distance of 200m. This scenario is shown in Fig.~\ref{system}. The background noise power spectral density is $-174$dBm/Hz. Channel gains are simulated using $L = 140.7 + 36.7 \log_{10}(d_{\mathrm{km}})$ dB as the path loss model, lognormal shadowing with $10$dB standard deviation, and Rayleigh fading \cite{3gpp}.

\subsection{Relaying for Interference Mitigation and Signal Enhancement}

%\begin{figure}[t]
%\center
%\input{myfig-11-v2.tex}
%\small
%\caption{Throughput improvement versus relaying capacity $C_0$ for the case where the source has three antennas ($s = 3$), and the relay and destination are equipped with two antennas each ($r = 2$ and $d = 2$) with no inter-cell interference ($t = 0$).} \label{fig:3_2_2_case}
%\end{figure}

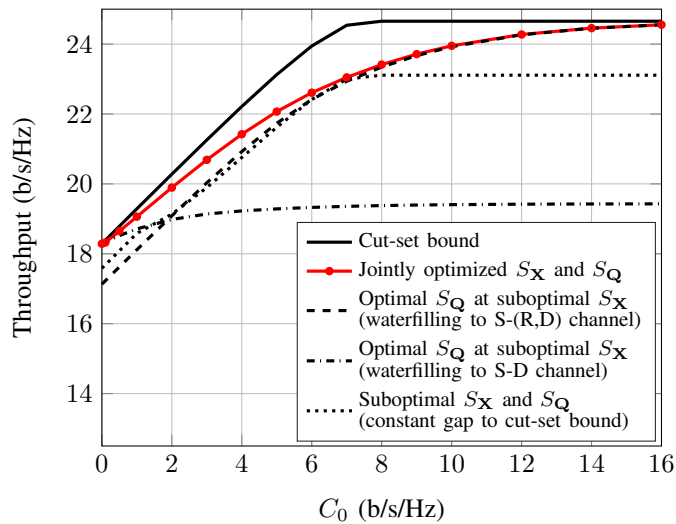
\begin{figure}[t]
\center
\newcommand{\LineWidth}{1.5}

\begin{tikzpicture}

\begin{axis}[
scale = 0.84,
width=\columnwidth,
height=0.78125\columnwidth,
at={(0in,0in)},
scale only axis,
%axis lines = middle,
%separate axis lines,
%axis x line*=bottom,
%every inner x axis line/.append style={line width=\LineWidth pt},
%every x tick label/.append style={font=\color{black}, line width=\LineWidth pt},
xmin=0,
xmax=16,
x label style={at={(axis description cs:0.5,-0.1)},anchor=north},
xlabel={$C_0$ (b/s/Hz)},
xmajorgrids,
%axis y line*=left,
%every inner y axis line/.append style={line width=\LineWidth pt},
%every y tick label/.append style={font=\color{black}, line width=\LineWidth pt},
ymin=12.5,
ymax=25,
y label style={at={(axis description cs:-0.1,.5)},rotate=0,anchor=south},
ylabel={Throughput (b/s/Hz)},
ymajorgrids,
axis background/.style={fill=white},
legend style={at={(0.99,0.01)},anchor=south east,legend cell align=left,align=left,draw=white!15!black,fill=white, font=\fontsize{8}{5}\selectfont}
]
\addplot [color=black,solid,line width=1.2pt]
  table[row sep=crcr]{%
0   18.2878\\
    0.1000   18.3878\\
    0.5000   18.7878\\
    1.0000   19.2878\\
    2.0000   20.2784\\
    3.0000   21.2573\\
    4.0000   22.2146\\
    5.0000   23.1274\\
    6.0000   23.9445\\
    7.0000   24.5396\\
    8.0000   24.6549\\
    9.0000   24.6549\\
   10.0000   24.6549\\
   12.0000   24.6549\\
   14.0000   24.6549\\
   16.0000   24.6549\\
   18.0000   24.6549\\
};
\addlegendentry{Cut-set bound};

\addplot [color=red,solid,line width=1.2pt,mark size=1.0pt,mark=o,mark options={solid}]
table[row sep=crcr]{
0   18.2878\\
    0.1000   18.3244\\
    0.5000   18.6480\\
    1.0000   19.0615\\
    2.0000   19.8949\\
    3.0000   20.6889\\
    4.0000   21.4210\\
    5.0000   22.0689\\
    6.0000   22.6094\\
    7.0000   23.0478\\
    8.0000   23.4144\\
    9.0000   23.7133\\
   10.0000   23.9507\\
   12.0000   24.2750\\
   14.0000   24.4567\\
   16.0000   24.5536\\
   18.0000   24.6037\\
};
\addlegendentry{Jointly optimized $S_{\bt X}$ and $S_{\bt Q}$};

\addplot [color=black,dashed,line width=1.2pt]
table[row sep=crcr]{%
0   17.1304\\
    0.1000   17.2294\\
    0.5000   17.6244\\
    1.0000   18.1159\\
    2.0000   19.0871\\
    3.0000   20.0314\\
    4.0000   20.9259\\
    5.0000   21.7355\\
    6.0000   22.4169\\
    7.0000   22.9368\\
    8.0000   23.3433\\
    9.0000   23.6698\\
   10.0000   23.9252\\
   12.0000   24.2671\\
   14.0000   24.4545\\
   16.0000   24.5530\\
   18.0000   24.6035\\
};
\addlegendentry{Optimal $S_{\bt Q}$ at suboptimal $S_{\bt X}$ \\ (waterfilling to S-(R,D) channel)};

\addplot [color=black,dashdotted,line width=1.2pt]
table[row sep=crcr]{
0   18.2878\\
    0.1000   18.3378\\
    0.5000   18.5209\\
    1.0000   18.7119\\
    2.0000   18.9823\\
    3.0000   19.1393\\
    4.0000   19.2263\\
    5.0000   19.2852\\
    6.0000   19.3277\\
    7.0000   19.3581\\
    8.0000   19.3797\\
    9.0000   19.3952\\
   10.0000   19.4061\\
   12.0000   19.4194\\
   14.0000   19.4261\\
   16.0000   19.4294\\
   18.0000   19.4311\\
};
\addlegendentry{Optimal $S_{\bt Q}$ at suboptimal $S_{\bt X}$ \\ (waterfilling to S-D channel)};

\addplot [color=black,dotted,line width=1.2pt]
table[row sep=crcr]{
0   17.5852\\
    0.1000   17.6852\\
    0.5000   18.0852\\
    1.0000   18.5852\\
    2.0000   19.1340\\
    3.0000   19.8948\\
    4.0000   20.7543\\
    5.0000   21.6206\\
    6.0000   22.4150\\
    7.0000   22.9990\\
    8.0000   23.1111\\
    9.0000   23.1111\\
   10.0000   23.1111\\
   12.0000   23.1111\\
   14.0000   23.1111\\
   16.0000   23.1111\\
   18.0000   23.1111\\
};
\addlegendentry{Suboptimal $S_{\bt X}$ and $S_{\bt Q}$ \\ (constant gap to cut-set bound)};

\end{axis}
\end{tikzpicture}
\small
\caption{Spectral efficiency versus relaying capacity $C_0$ for the case where the source has three antennas ($s = 3$), and the relay and destination are each equipped with two antennas ($r = 2$ and $d = 2$) with no inter-cell interference ($t = 0$).} \label{fig:3_2_2_case}
\end{figure}

%\begin{figure}[t]
%\center
%\input{myfig-22-v2.tex}
%\small
%\caption{Throughput improvement versus relaying capacity
%$C_0$ for the case where the source has two antennas ($s = 2$), and the relay and destination are each equipped with three antennas ($r = 3$ and $d = 3$) with no inter-cell interference ($t = 0$).}
%\label{fig:2_3_3_case}
%\end{figure}

\begin{figure}[t]
\center
\newcommand{\LineWidth}{1.5}

\begin{tikzpicture}

\begin{axis}[%
scale = 0.84,
width=\columnwidth,
height=0.78125\columnwidth,
at={(0in,0in)},
scale only axis,
separate axis lines,
%axis x line=bottom,
%every inner x axis line/.append style={line width= \LineWidth pt},
%every x tick label/.append style={font=\color{black}, line width=\LineWidth pt},
xmin=0,
xmax=8,
x label style={at={(axis description cs:0.5,-0.1)},anchor=north},
xlabel={$C_0$ (b/s/Hz)},
xmajorgrids,
%axis y line=left,
%every inner y axis line/.append style={line width= \LineWidth pt},
%every y tick label/.append style={font=\color{black}, line width=\LineWidth pt},
ymin=15,
ymax=21.25,
y label style={at={(axis description cs:-0.1,.5)},rotate=0,anchor=south},
ylabel={Throughput (b/s/Hz)},
ymajorgrids,
axis background/.style={fill=white},
legend style={at={(0.99,0.01)},anchor=south east,legend cell align=left,align=left,draw=white!15!black,fill=white, font=\fontsize{8}{5}\selectfont}
]
\addplot [color=black,solid,line width=1.2pt]
  table[row sep=crcr]{%
0   18.2878\\
    0.1000   18.3878\\
    0.3000   18.5878\\
    0.5000   18.7878\\
    1.0000   19.2878\\
    1.5000   19.7878\\
    2.0000   20.1061\\
    3.0000   20.1061\\
    4.0000   20.1061\\
    5.0000   20.1061\\
    6.0000   20.1061\\
    7.0000   20.1061\\
    8.0000   20.1061\\
};
\addlegendentry{Cut-set bound};

\addplot [color=red,solid,line width=1.2pt,mark size=1.0pt,mark=o,mark options={solid}]
  table[row sep=crcr]{%
0   18.2878\\
    0.1000   18.3540\\
    0.3000   18.4818\\
    0.5000   18.6031\\
    1.0000   18.8766\\
    1.5000   19.1071\\
    2.0000   19.2960\\
    3.0000   19.5629\\
    4.0000   19.7196\\
    5.0000   19.8275\\
    6.0000   19.9063\\
    7.0000   19.9634\\
	8   20.004\\
};
\addlegendentry{Jointly optimized $S_{\bt X}$ and $S_{\bt Q}$};

\addplot [color=black,dashed,line width=1.2pt]
  table[row sep=crcr]{%
0   18.2878\\
    0.1000   18.3296\\
    0.3000   18.4104\\
    0.5000   18.4877\\
    1.0000   18.6662\\
    1.5000   18.8252\\
    2.0000   18.9667\\
    3.0000   19.2044\\
    4.0000   19.3923\\
    5.0000   19.5408\\
    6.0000   19.6583\\
    7.0000   19.7512\\
   8   19.8247\\
};
\addlegendentry{Optimized $S_{\bt X}$, and $S_{\bt Q} = q \bt I_r$};

\addplot [color=black,dotted,line width=1.2pt]
  table[row sep=crcr]{%
0   17.3596\\
    0.3000   17.6596\\
    0.5000   17.8596\\
    1.5000   18.8596\\
    2.0000   19.1779\\
    3.0000   19.2693\\
    5.0000   19.2693\\
    7.0000   19.2693\\
   8   19.2693\\
};
\addlegendentry{Suboptimal $S_{\bt X}$ and $S_{\bt Q}$ \\ (constant gap to cut-set bound)};

\end{axis}
\end{tikzpicture}%
\small
\caption{Spectral efficiency versus relaying capacity
$C_0$ for the case where the source has two antennas
($s = 2$), and the relay and destination are each equipped with three antennas 
($r = 3$ and $d = 3$) with no inter-cell interference ($t = 0$).}
\label{fig:2_3_3_case}
\end{figure}
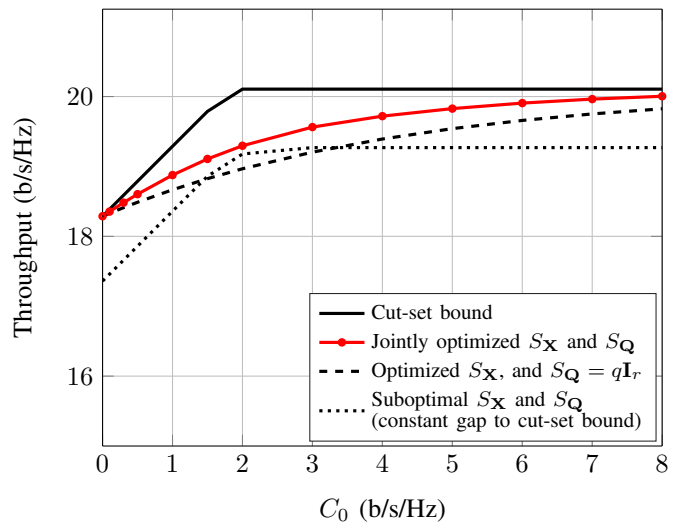

%\begin{figure}[h]
%\center
%\input{myfig-333-v2.tex}
%\small
%\caption{Throughput improvement relaying capacity $C_0$ for the case where the source has two antennas ($s=2$), and the relay and destination are each equipped with three antennas ($r = 3$ and $d = 3$) with four inter-cell interference sources ($t = 4$). }
%\label{fig:2_3_3_4_case}
%\end{figure}

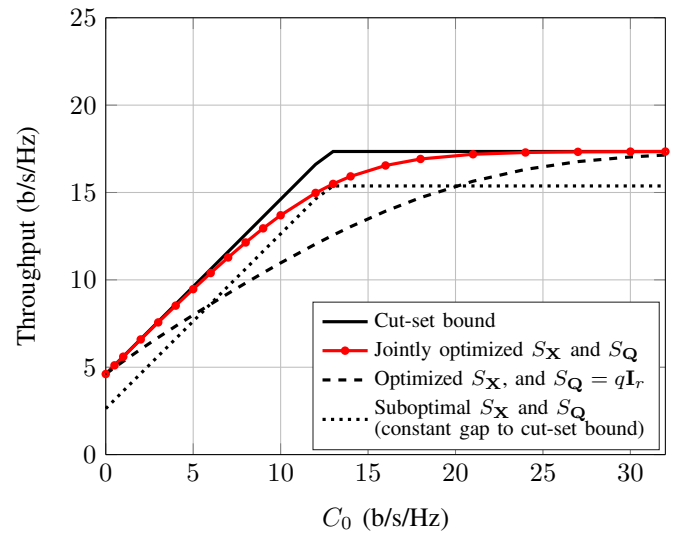
\begin{figure}[h]
\center
\newcommand{\LineWidth}{1.5}

\begin{tikzpicture}

\begin{axis}[
scale = 0.84,
width=\columnwidth,
height=0.78125\columnwidth,
at={(0in,0in)},
scale only axis,
separate axis lines,
%axis x line=bottom,
%every inner x axis line/.append style={line width= \LineWidth pt},
%every x tick label/.append style={font=\color{black}, line width=\LineWidth pt},
xmin=0,
xmax=32,
x label style={at={(axis description cs:0.5,-0.1)},anchor=north},
xlabel={$C_0$ (b/s/Hz)},
xmajorgrids,
%axis y line=left,
%every inner y axis line/.append style={black,line width=1.5pt},
%every y tick label/.append style={font=\color{black}},
ymin=0,
ymax=25,
y label style={at={(axis description cs:-0.1,.5)},rotate=0,anchor=south},
ylabel={Throughput (b/s/Hz)},
ymajorgrids,
axis background/.style={fill=white},
legend style={at={(0.99,0.01)},anchor=south east,legend cell align=left,align=left,draw=white!15!black,fill=white, font=\fontsize{8}{5}\selectfont}
]
\addplot [color=black,solid,line width=1.2pt]
  table[row sep=crcr]{%
     0    4.6109\\
    0.5000    5.1109\\
    1.0000    5.6109\\
    2.0000    6.6109\\
    3.0000    7.6109\\
    4.0000    8.6109\\
    5.0000    9.6109\\
    6.0000   10.6109\\
    7.0000   11.6109\\
    8.0000   12.6109\\
    9.0000   13.6109\\
   10.0000   14.6109\\
   12.0000   16.6109\\
   13.0000   17.3470\\
   14.0000   17.3470\\
   16.0000   17.3470\\
   18.0000   17.3470\\
   21.0000   17.3470\\
   24.0000   17.3470\\
   27.0000   17.3470\\
   30.0000   17.3470\\
   32         17.3470\\
};
\addlegendentry{Cut-set bound};

\addplot [color=red,solid,line width=1.2pt,mark size=1.0pt,mark=o,mark options={solid}]
  table[row sep=crcr]{%
0    4.6109\\
    0.5000    5.1082\\
    1.0000    5.6044\\
    2.0000    6.5915\\
    3.0000    7.5661\\
    4.0000    8.5254\\
    5.0000    9.4687\\
    6.0000   10.3906\\
    7.0000   11.2838\\
    8.0000   12.1395\\
    9.0000   12.9474\\
   10.0000   13.6963\\
   12.0000   14.9757\\
   13.0000   15.4915\\
   14.0000   15.9222\\
   16.0000   16.5466\\
   18.0000   16.9188\\
   21.0000   17.1881\\
   24.0000   17.2898\\
   27.0000   17.3266\\
   30.0000   17.3398\\
   32   17.3434\\
};
\addlegendentry{Jointly optimized $S_{\bt X}$ and $S_{\bt Q}$};

\addplot [color=black,dashed,line width=1.2pt]
  table[row sep=crcr]{%
0    4.6109\\
    0.5000    4.9780\\
    1.0000    5.3365\\
    2.0000    6.0313\\
    3.0000    6.7017\\
    4.0000    7.3526\\
    5.0000    7.9880\\
    6.0000    8.6103\\
    7.0000    9.2204\\
    8.0000    9.8178\\
    9.0000   10.4016\\
   10.0000   10.9697\\
   12.0000   12.0505\\ 
   13.0000   12.5586\\
   14.0000   13.0425\\
   16.0000   13.9309\\
   18.0000   14.7057\\
   21.0000   15.6452\\
   24.0000   16.3224\\
   27.0000   16.7677\\
   30.0000   17.0352\\
32   17.1445\\
};
\addlegendentry{Optimized $S_{\bt X}$, and $S_{\bt Q} = q \bt I_r$};

\addplot [color=black,dotted,line width=1.2pt]
  table[row sep=crcr]{%
0    2.6323\\
    0.5000    3.1323\\
    1.0000    3.6387\\
    2.0000    4.6387\\
    3.0000    5.6387\\
    4.0000    6.6323\\
    5.0000    7.6323\\
    6.0000    8.6323\\
    7.0000    9.6387\\
    8.0000   10.6323\\
    9.0000   11.6387\\
   10.0000   12.6323\\
   12.0000   14.6323\\
   13.0000   15.3747\\
   14.0000   15.3747\\
   16.0000   15.3747\\
   18.0000   15.3747\\
   21.0000   15.3747\\
   24.0000   15.3747\\
   27.0000   15.3747\\
   30.0000   15.3747\\
32.0000   15.3747\\
};
\addlegendentry{Suboptimal $S_{\bt X}$ and $S_{\bt Q}$ \\ (constant gap to cut-set bound)};

\end{axis}
\end{tikzpicture}%
\small
\caption{Spectral efficiency versus relaying capacity $C_0$ for the case where the source has two antennas ($s=2$), and the relay and destination are each equipped with three antennas ($r = 3$ and $d = 3$) with four inter-cell interference sources ($t = 4$). }
\label{fig:2_3_3_4_case}
\end{figure}

Fig.~\ref{fig:3_2_2_case} illustrates the effect of antenna pooling for enhancing the intended signal when there is no interference, so the noises at relay and destination are independent. It shows the rate improvement by relaying in a channel with $(s, d, r, t) = (3, 2, 2, 0)$. Without the relay, DoF is limited by the number of antennas at the destination, which is $DoF_D = 2$. Adding $r = 2$ extra antennas to the destination improves DoF by $\Delta DoF = 1$. Also, this channel has $1$ asymptotically deterministic component, and as shown in Fig.~\ref{fig:3_2_2_case}, at $C_0 = 1$ b/s/Hz, the improvement in throughput by the optimized compress-and-forward is around $0.74$ b/s/Hz. Here, antenna pooling improves the overall throughput considerably.

Fig.~\ref{fig:3_2_2_case} also demonstrates the importance of optimizing the input covariance matrix. We plot the achievable rates for two suboptimal choices of input covariance: water-filling covariance of the source-to-destination and source-to-relay-and-destination point-to-point channels. For these fixed input covariance matrices, we optimize the quantization noise covariance. Depending on the value of $C_0$, each of them can be strictly suboptimal. For small values of $C_0$, steering the beam toward destination is close to optimal. However, as $C_0$ increases, this $S_{\bt X}$ fails to achieve benefits of antenna pooling. For large values of $C_0$, steering the beam toward both relay and destination is close to optimal, but a gap exits at small $C_0$. Moreover, in this scenario, joint optimization of input and quantization noise outperforms the suboptimal evaluation of achievable rate for constant-gap capacity characterization by up to 6.5\%. %The joint optimization algorithm outperforms this suboptimal scheme at small $C_0$ by at most 4\%.

Fig.~{\ref{fig:2_3_3_case} presents the results for a channel with $\left(s, d, r, t\right) = \left(2, 3, 3, 0\right)$. Without the relay, DoF is constrained by the number of BS antennas, not the number of antennas at the destination. Therefore, relaying does not increase the overall DoF. Also, as predicted by Theorem \ref{thm_finite_infinite}, the improvement in the overall rate is not notable. In this case, with $C_0 = 7$ b/s/Hz, the throughput improvement is only around $1.7$ b/s/Hz. 
%For smaller values of $C_0$ the optimized CF rate also has a considerable gap to the cut-set bound.  However, if this experiment is performed in presence of noise correlation, MIMO relaying is able to improve throughput drastically, through interference rejection.

Fig.~\ref{fig:2_3_3_4_case} considers the same setup as Fig.~\ref{fig:2_3_3_case}, except that now four interfering single-antenna BSs are added and we have $\left(s, d, r, t\right) = \left(2, 3, 3, 4\right)$. Due to interference, the rate at $C_0 = 0$ is low, but the optimized use of the relaying link improves the throughput significantly. This is because, due to the common intercell interference, now the noises at the relay and at the destination are correlated. This noise correlation makes relaying effective for improving the rate. The maximum possible rate improvement is around $12.7$ b/s/Hz, which is achieved at $C_0 = 24$ b/s/Hz. Here, the DoF improvement with infinite $C_0$ is $\Delta DoF = 2$ and the channel has $2$ asymptotically deterministic components. At around $C_0 = 10$ b/s/Hz, the improvement in throughput is around $9.1$ b/s/Hz. In this scenario, the improvement in rate by joint optimization of input and quantization noise covariance matrices over the suboptimal evaluation of rate for constant-gap capacity characterization is around 75\% at small $C_0$'s and 12.8\% at large $C_0$'s, illustrating the importance of optimizing quantization at low $C_0$'s. It is worth noting that the overall throughput at large $C_0$ in Fig.~\ref{fig:2_3_3_4_case} is close to the achievable rate of the $C_0 = 0$ scenario in Figs.~\ref{fig:3_2_2_case} and \ref{fig:2_3_3_case}, illustrating the almost complete interference rejection capability of relaying.

Figs.~\ref{fig:2_3_3_case} and \ref{fig:2_3_3_4_case} also demonstrate the
importance of optimizing the quantization noise covariance matrix. The achievable rate for a simple suboptimal choice of $S_{\bt Q} = q \bt I_r$ is included, where $q$ is set to satisfy the relaying rate constraint with equality, for the optimized $S_{\bt X}$ obtained from Algorithm \ref{algo}. This simple choice of $S_{\bt Q}$ results in a strictly suboptimal performance as shown in Figs.~\ref{fig:2_3_3_case} and \ref{fig:2_3_3_4_case}. In the scenario of Fig.~\ref{fig:2_3_3_4_case}, this simple choice of $S_{\bt Q}$ results in a DoF loss as shown by Theorem \ref{thm_sub_opt_dof}.

%It is worth noting that the slope of the optimized compress-and-forward rate remains close to 1 for $C_0$ around $50\mathrm{Mbps}$ in Fig.~\ref{fig:3_2_2_case} whereas it remains close to 1 for $C_0$ around $100\mathrm{Mbps}$ in Fig.~\ref{fig:2_3_3_4_case}. This happens because, according to lemma \ref{lemma_det}, in the scenario of Fig.~\ref{fig:3_2_2_case} the relay channel has only one deterministic component whereas there are two deterministic components in the scenario of Fig.~\ref{fig:2_3_3_4_case}.

\subsection{Effect of Numbers of Antennas on the Slope}

%\begin{figure*}[t]
%\begin{center}
%
\caption{Average slope of $\bar R_{CF}(C_0)$ at $\bar C_{0,i}$, i.e., $1 - \frac{1}{\lambda_i}$, for $i = 1, \dots, 6$ as numbers of antennas at the destination ($d$) and at the relay ($r$) vary, with $s =5$ antennas at the source and $t = 18$ single-antenna interference sources. Observe that the average slope at $\bar C_{0,i}$ is close to 1 when $r + d > t + i - 1$ and $d < s + t - i + 1$ for $i = 1, \dots, 5$, and is close to zero for $i = 6$.} \label{fig_slope}
\end{center}
\end{figure*}

We illustrate the effect of number of antennas at the relay ($r$) and at the destination ($d$) on the slope of compress-and-forward rate versus relaying link capacity $\bar R_{CF}(C_0)$. In this example, the source is equipped with five antenna and there are 18 interfering BSs each equipped with one antenna, i.e., $(s, t) = (5, 18)$. Fig.~\ref{fig_slope} shows the value of $1 - \frac{1}{\lambda_i}$ for $i = 1, \dots, 6$, which is the slope of $\bar R_{CF}(C_0)$ at $\bar C_i$ as in (\ref{crit_slope}), averaged over 100 realizations of the channel and fixed transmit covariance $S_{\bt X}$.

Since the source has $s = 5$ antennas, by Theorem \ref{prop_rev_deg}, the $6^{\mathrm{th}}$ element of $C_R \bt Y_R$ is always reversely degraded and its CSINR is zero, i.e., $\lambda_6 = 1$. The optimal quantization does not quantize this element, because it cannot further improve the rate. In simulation results shown in Fig. \subref*{fig_slope_6}, the slope of rate curve at $\bar C_6$ is almost zero.

For $i = 1, \dots, 5$, by (\ref{det_comps}) in Theorem \ref{thm_det_comp}, when $r$ and $d$ are such that both $r + d > 17 + i$ and $d < 24 - i$, the $i^{\mathrm{th}}$ element of $C_R \bt Y_R$ is asymptotically deterministic. Hence, the value of $1 - \frac{1}{\lambda_i}$ approaches 1 as the power of noise goes to zero. When $d \ge 24 - i$ the $i^{\mathrm{th}}$ element of $C_R \bt Y_R$ is asymptotically reversely degraded, and $1 - \frac{1}{\lambda_i}$ does not approach 1. Also, when $r + d \le 17 + i$, the 18-dimensional interference signal makes both $S_{ {\bt Y}_R | {\bt Y}_D \bt X}$ and $S_{ {\bt Y}_R | {\bt Y}_D}$ full-rank, and keeps the slope $1 - \frac{1}{\lambda_i}$ away from 1. These are indeed observed by numerical simulations in Figs. \ref{fig_slope}\subref{fig_slope_1}-\subref{fig_slope_5}.

\section{Conclusion}
\label{section_con}

This paper considers MIMO relaying for interference mitigation and signal enhancement in a wireless communication network. %The optimized vector compress-and-forward enables efficient pooling of relay's antennas with the receiver over a digital link of finite-capacity. 
Joint optimization of transmit and quantization noise covariance matrices in compress-and-forward scheme enables the receiver to efficiently utilize extra spatial dimensions from a relay node. This paper shows that this scheme achieves the capacity of the channel to within a constant gap that only depends on the numbers of antennas. %As a suboptimal, but simpler scheme, one can consider fixing the transmit covariance matrix to the solution of water-filling against the source-to-relay-and-destination point-to-point channel, and only optimizing quantization at the relay for exploiting the available $C_0$. Performance of this scheme is close to the optimized compress-and-forward rate in the numerical simulations.

Optimizing the relay's quantization noise covariance matrix is crucial for efficient interference mitigation. This is the key step in characterizing the capacity of the MIMO relay channel to within a constant gap. It further reveals the asymptotically deterministic and reversely degraded components of the channel. When the channel has an asymptotically deterministic component, the slope of the optimized compress-and-forward rate $\bar{R}_{CF}(C_0)$ at small $C_0$ approaches its maximum of 1 at high SNR and INR. Optimizing relay's quantization enables achieving the maximum possible DoF of the channel through distributed zero-forcing of interference.

Antenna pooling is most efficient when the number of antennas at the source $s$, at the destination $d$, at the relay $r$, and the dimension of interference $t$ satisfy both $r + d > t$ and $d < s + t$; otherwise, it does not improve the overall DoF at $C_0 = \infty$, and the slope of $\bar R_{CF}(C_0)$ at small $C_0$ does not approach the maximum of 1 at high SNR and INR. Typically, the number of antennas at a user device is small and the latter condition easily holds. The former condition points to the benefit of deploying a large number of antennas at the relay node, which enables distributed zero-forcing of the interference. %This is under the assumption that in the network SNR and INR are proportional. To have efficient relaying in networks with heterogeneous interference powers, the relay should be equipped with enough antennas for overcoming, at least, the dimension of dominant interference.

%We observed that fixing the input covariance matrix to water-filling solution of source-to-relay-and-destination point-to-point channel, and, then, optimizing the quantization scheme at the relay for the given value of $C_0$ achieves most of potentialities of antenna pooling in improving throughput.

%A scenario where interference, regardless of its power, is treated as a part of noise. Treating interference as noise is simple and is of practical importance.

%In the compress-and-forward lower bound having an asymptotic slope of 1 for $R_{CF}(C_0)$ at finite $C_0$ is equivalent to having DoF gain by relaying at $C_0 = \infty$. Our analysis shows that for having efficient relaying the sum of numbers of antennas at the relay and destination nodes must be as large as dimension of interference. 

% if have a single appendix:
%\appendix[Proof of the Zonklar Equations]
% or
%\appendix  % for no appendix heading
% do not use \section anymore after \appendix, only \section*
% is possibly needed

% use appendices with more than one appendix
% then use \section to start each appendix
% you must declare a \section before using any
% \subsection or using \label (\appendices by itself
% starts a section numbered zero.)
%

\appendices
%\section{Two Lemmas}
\section{}
\label{app_lemmas}

The following lemma characterizes the scaling of eigenvalues of $S_{\bt Y_R \mid \bt Y_D}$ and $S_{\bt Y_R \mid \bt Y_D, \bt X}$ with the background noise power, as it goes to zero. It is used for counting the number of asymptotic deterministic components and for the DoF analysis in Appendices \ref{app_prf_thm_det_comp} to \ref{app_prf_thm_opt_dof}.
\begin{lemma}
At the limit of high SNR and INR, i.e., as $\sigma^2 \to 0$, we have
\begin{align}
\lambda_i(S_{\bt Y_R \mid \bt Y_D}) &= \lambda_i(S_{\bar{\bt Y}_R \mid \bar{\bt Y}_D}) 
+ a_i \sigma^2 + O(\sigma^4), \\
\lambda_i(S_{\bt Y_R \mid \bt Y_D, \bt X}) &= \lambda_i(S_{\bar{\bt Y}_R \mid \bar{\bt Y}_D, \bt X}) + a'_i \sigma^2 + O(\sigma^4),  \label{lambda_scaling_1}
\end{align}
for positive $a_i$'s and $a'_i$'s. Here, $\bar{\bt Y}_R$ and $\bar{\bt Y}_D$ are as defined in (\ref{r_nonoise})-(\ref{d_nonoise}).
\label{lemma_limit}
\end{lemma}

\begin{IEEEproof} Define $L \triangleq \left[\begin{matrix} H_{SD}S_{\bt X}^{\frac{1}{2}} \ \ \  H_{TD}S_{\bt X_T}^{\frac{1}{2}} \end{matrix}\right]$ and $M \triangleq \left[\begin{matrix} H_{SR}S_{\bt X}^{\frac{1}{2}} \ \ \  H_{TR}S_{\bt X_T}^{\frac{1}{2}} \end{matrix}\right]$. We have
\begin{align}
&S_{{\bt Y}_R \mid {\bt Y}_D}  \nonumber \\ & \quad \stackrel{(a)}{=} S_{\bar{\bt Y}_R} + \sigma^2 \bt I_r - M L^\dagger \left(LL^\dagger + \sigma^2 \bt I_d \right)^{-1} L M^\dagger \nonumber \\ & \quad \stackrel{(b)}{=} S_{\bar{\bt Y}_R} + \sigma^2 \bt I_r - M V D^\dagger \left(D D^\dagger + \sigma^2 \bt I_d \right)^{-1} D V^\dagger M^\dagger \nonumber \\ & \quad \stackrel{(c)}{=} S_{\bar{\bt Y}_R} + \sigma^2 \bt I_r - M V D^\dagger \nonumber \\ & \qquad \qquad \qquad \quad \quad \ \left(\sum_{n = 0}^{\infty} (-\sigma^2)^n(D D^\dagger)^{-(n+1)}\right)  D V^\dagger M^\dagger \nonumber \\ & \quad = S_{\bar{\bt Y}_R \mid \bar{\bt Y}_D} + \sigma^2 \bt I_r - M L^\dagger  \left(\sum_{n=1}^{\infty}(-\sigma^2)^n S_{\bar{\bt Y}_D}^{-(n+1)}\right) L M^\dagger \nonumber \\ & \quad \triangleq S_{\bar{\bt Y}_R \mid \bar{\bt Y}_D} + \sum_{n=1}^{\infty}(-1)^{n+1}\sigma^{2n} S_{n},
\label{taylor_s}
\end{align}
where, $(a)$ follows by the Schur complement formula, $(b)$ follows by the singular value decomposition (SVD) $L = U D V^{\dagger}$, and $(c)$ follows by the Taylor expansion of each diagonal element.

Similar to the above, by taking SVD of $H_{TD} S_{\bt X_T}^{\frac{1}{2}}$ one can write
\begin{align}
& S_{{\bt Y}_R \mid {\bt Y}_D, \bt X} = S_{\bar{\bt Y}_R \mid \bar{\bt Y}_D, \bt X} + \sum_{n=1}^{\infty}(-1)^{n+1}\sigma^{2n} S'_{n}.
\label{taylor_s'}
\end{align}
All $S_n$'s and $S'_n$'s are positive semidefinite, and both $S_1$ and $S'_1$ are full-rank.

Since $S_{{\bt Y}_R \mid {\bt Y}_D}$ and $S_{{\bt Y}_R \mid {\bt Y}_D, \bt X}$ have convergent series in $\sigma^2$ with Hermitian $S_n$'s and $S'_n$'s, the eigenvalues $\lambda_i(S_{{\bt Y}_R \mid {\bt Y}_D})$ and $\lambda_i(S_{{\bt Y}_R \mid {\bt Y}_D, \bt X})$ have convergent series in $\sigma^2$ as well \cite[Chapter~1]{perturbation}
\begin{align}
&\lambda_i(S_{{\bt Y}_R \mid {\bt Y}_D}) = \sum_{n=0}^{\infty} a_{i,n}\sigma^{2n},
\label{taylor_lambda_s}
\end{align}
and
\begin{align}
&\lambda_i(S_{{\bt Y}_R \mid {\bt Y}_D, \bt X}) = \sum_{n=0}^{\infty}a'_{i,n}\sigma^{2n}.
\label{taylor_lambda_s'}
\end{align}

Using Weyl's inequality \cite[Theorem 4.3.1]{horn} over the summations in (\ref{taylor_s}) and (\ref{taylor_s'}) inductively, we have
\begin{multline}
\lambda_i(S_{\bar{\bt Y}_R \mid \bar{\bt Y}_D}) + \sum_{n=1}^{\infty}(-1)^{n+1}\sigma^{2n} \lambda_r(S_{n}) \leq \lambda_i(S_{{\bt Y}_R \mid {\bt Y}_D}) \\ \leq \lambda_i(S_{\bar{\bt Y}_R \mid \bar{\bt Y}_D}) + \sum_{n=1}^{\infty}(-1)^{n+1}\sigma^{2n} \lambda_1(S_{n}),
\label{weyl_s}
\end{multline}
and
\begin{multline}
\lambda_i(S_{\bar{\bt Y}_R \mid \bar{\bt Y}_D, \bt X}) + \sum_{n=1}^{\infty}(-1)^{n+1}\sigma^{2n} \lambda_r(S'_{n}) \leq \lambda_i(S_{{\bt Y}_R \mid {\bt Y}_D, \bt X}) \\ \leq \lambda_i(S_{\bar{\bt Y}_R \mid \bar{\bt Y}_D, \bt X}) + \sum_{n=1}^{\infty}(-1)^{n+1}\sigma^{2n} \lambda_1(S'_{n}),
\label{weyl_s'}
\end{multline}
By the squeeze theorem, $a_{i,0} = \lambda_i(S_{\bar{\bt Y}_R \mid \bar{\bt Y}_D})$ and $a'_{i,0} =\lambda_i(S_{\bar{\bt Y}_R \mid \bar{\bt Y}_D, \bt X})$. Moreover, $a_{i,1}$ and $a'_{i,1}$ are positive, because both $S_1$ and $S'_1$ are positive definite matrices.
\end{IEEEproof}

\section{Proof of Theorem \ref{thm_det_comp}}
\label{app_prf_thm_det_comp}

Before counting the number of asymptotic deterministic components, we characterize the rank of the two conditional covariance matrices in the following lemma.

\begin{lemma}
Consider $\bar{\bt Y}_R$ and $\bar{\bt Y}_D$ as defined in (\ref{r_nonoise})-(\ref{d_nonoise}). We have $\mathrm{rank}(S_{\bar{\bt Y}_R \mid \bar{\bt Y}_D}) = \min \left( r , (s^\mathrm{r}+t-d)^+ \right)$ and $\mathrm{rank}(S_{\bar{\bt Y}_R \mid \bar{\bt Y}_D, \bt X})= \min \left( r , (t-d)^+ \right)$, almost surely.
\label{lemma_rank}
\end{lemma}
\begin{IEEEproof} Since $\mathrm{rank}(S_{\bt X}) = s^\mathrm{r}$ and $\mathrm{rank}(S_{\bt X_T}) = t$, we have
\begin{eqnarray}
& \mathrm{rank}(S_{ \bar{\bt Y}_R, \bar{\bt Y}_D}) &= \mathrm{min}( r+d , s^\mathrm{r}+t), \label{rank_rd} \\
& \mathrm{rank}(S_{ \bar{\bt Y}_D}) &= \mathrm{min}(d , s^\mathrm{r}+t), \label{rank_d}\\
& \mathrm{rank}(S_{ \bar{\bt Y}_R , \bar{\bt Y}_D, \bt X}) &= s^\mathrm{r} + \mathrm{min}(r+d , t), \label{rank_rdx}\\
& \mathrm{rank}(S_{ \bar{\bt Y}_D, \bt X}) &= s^\mathrm{r} + \mathrm{min}(d , t). \label{rank_dx}
\end{eqnarray}

By rank additivity of generalized Schur complement \cite[Theorem 1]{gsc}
\begin{align}
\mathrm{rank}(S_{ \bar{\bt Y}_R \mid \bar{\bt Y}_D}) \leq \mathrm{rank}(S_{ \bar{\bt Y}_R, \bar{\bt Y}_D}) - \mathrm{rank}(S_{\bar{\bt Y}_D}),
\label{rank_add2}
\end{align}
with equality if the null space of $S_{ \bar{\bt Y}_D}$ is a subset of the null space of $S_{ \bar{\bt Y}_R, \bar{\bt Y}_D}^{(1,2)}$. When $d \leq s^\mathrm{r} + t$, $S_{ \bar{\bt Y}_D}$ is full-rank and has a trivial null space. When $d \geq s^\mathrm{r} + t$, the right-hand-side of (\ref{rank_add2}) is zero. Therefore, equality always holds in (\ref{rank_add2}), and
\begin{align} 
\mathrm{rank}(S_{ \bar{\bt Y}_R \mid \bar{\bt Y}_D}) = \min \left( r , (s+t-d)^+ \right).
\end{align}
Similarly, we have
\begin{align}
\mathrm{rank}(S_{ \bar{\bt Y}_R \mid \bar{\bt Y}_D , \bt X}) \leq \mathrm{rank}(S_{ \bar{\bt Y}_R, \bar{\bt Y}_D, \bt X}) - \mathrm{rank}(S_{\bar{\bt Y}_D, \bt X}),
\label{rank_add1}
\end{align}
with equality if the null space of $S_{ \bar{\bt Y}_D, \bt X}$ is a subset of the null space of $S_{ \bar{\bt Y}_R, \bar{\bt Y}_D, \bt X}^{(1,2)}$. When $d \leq t$, the null space of
\begin{align*}
S_{ \bar{\bt Y}_D, \bt X} = \left[\begin{matrix}H_{SD} S_{\bt X}H_{SD}^\dagger+H_{TD} S_{\bt X_T}H_{TD}^\dagger  &  H_{SD}S_{\bt X} \\ S_{\bt X}H_{SD}^\dagger  &  S_{\bt X} \end{matrix}\right]
\end{align*}
is $s-s^\mathrm{r}$ dimensional and is caused by the rank deficiency of $S_{\bt X}$. Hence, it is a subset of the null space of
\begin{align*}
S_{ \bar{\bt Y}_R, \bar{\bt Y}_D, \bt X}^{(1,2)} = \left[\begin{matrix}H_{SR} S_{\bt X}H_{SD}^\dagger+H_{TR} S_{\bt X_T}H_{TD}^\dagger  &  H_{SR}S_{\bt X} \end{matrix}\right].
\end{align*}
When $d > t$, the right-hand-side of (\ref{rank_add1}) is zero. Therefore, we always have equality in (\ref{rank_add1}), and
\begin{align}
\mathrm{rank}(S_{ \bar{\bt Y}_R \mid \bar{\bt Y}_D, \bt X}) = \min \left(r, \left(t -d\right)^+ \right).
\end{align}
\end{IEEEproof}

Now, we proceed to prove Theorem \ref{thm_det_comp}.

\begin{IEEEproof}
The rank of the conditional covariance matrices $S_{\bar{\bt Y}_R \mid \bar{\bt Y}_D}$ and $S_{ \bar{\bt Y}_R \mid \bar{\bt Y}_D, \bt X}$ in (33) and (34) are characterized in Lemma \ref{lemma_rank}. Here, we argue that the largest $i$ for which we have
\begin{align}
\lim_{\sigma^2 \to 0} \lambda_i(S_{\bt Y_R \mid \bt Y_D}S_{ \bt Y_R \mid \bt Y_D, \bt X}^{-1}) = \infty
\end{align}
is the difference of the two ranks, $r' - r''$. To relate the limit of the generalized eigenvalues to ranks of conditional covariances, we need to use bounds
\begin{align}
\frac{\lambda_{r'' + i}(S_{\bt Y_R \mid \bt Y_D})}{\lambda_{r'' + 1}(S_{ \bt Y_R \mid \bt Y_D, \bt X})} \leq \lambda_i(S_{\bt Y_R \mid \bt Y_D}S_{ \bt Y_R \mid \bt Y_D, \bt X}^{-1}),
\label{eig_lbounds}
\end{align}
\begin{align}
\lambda_i(S_{\bt Y_R \mid \bt Y_D}S_{ \bt Y_R \mid \bt Y_D, \bt X}^{-1}) \leq \frac{\lambda_1(S_{\bt Y_R \mid \bt Y_D})}{\lambda_{r - (i - 1)}(S_{ \bt Y_R \mid \bt Y_D, \bt X})},
\label{eig_ubound1}
\end{align}
and
\begin{align}
\lambda_i(S_{\bt Y_R \mid \bt Y_D}S_{ \bt Y_R \mid \bt Y_D, \bt X}^{-1}) \leq \frac{\lambda_i(S_{\bt Y_R \mid \bt Y_D})}{\lambda_r(S_{ \bt Y_R \mid \bt Y_D, \bt X})},
\label{eig_ubound2}
\end{align}
due to \cite[Corollary 2.5]{eigvalbounds}.

For $i \leq r'-r''$, we have $\lambda_{r''+1}(S_{\bar{\bt Y}_R \mid \bar{\bt Y}_D, \bt X}) = 0$ and $\lambda_{r''+i}(S_{\bar{\bt Y}_R \mid \bar{\bt Y}_D}) > 0$. In this case, the $i^\mathrm{th}$ element is asymptotically deterministic, but not asymptotically reversely degraded. By taking limit from both sides of (\ref{eig_lbounds}) and using Lemma \ref{lemma_limit}, we have
\begin{multline}
\lim_{\sigma^2 \to 0} \lambda_i(S_{\bt Y_R \mid \bt Y_D}S_{ \bt Y_R \mid \bt Y_D, \bt X}^{-1}) \geq \lim_{\sigma^2 \to 0} \frac{\lambda_{r''+i}(S_{{\bt Y}_R \mid {\bt Y}_D})}{\lambda_{r''+1}(S_{ {\bt Y}_R \mid {\bt Y}_D, \bt X})} \\ = \lim_{\sigma^2 \to 0} \frac{\lambda_{r''+i}(S_{\bar{\bt Y}_R \mid \bar{\bt Y}_D})+a_{r''+i}\sigma^2}{a'_{r''+1}\sigma^2} = \infty.
\end{multline}

For $i > r'-r'' = \min \left(r, s^\mathrm{r}, \left(r+d-t\right)^+, \left(s^\mathrm{r}+t-d\right)^+ \right)$, the generalized eigenvalues remain finite. To see this, we should consider three possible cases.

1) When $i > s^\mathrm{r}$, by Theorem \ref{prop_rev_deg}, the $i^\mathrm{th}$ element is reversely degraded and we have $\lambda_i = 1$ at all values of $\sigma^2$.

2) When $i > r + d - t$, using (\ref{r'})-(\ref{r''}) it is easy to see that $r'' > r-i$ and $r' > 0$. Therefore, both $\lambda_{r-i+1}(S_{ \bar{\bt Y}_R \mid \bar{\bt Y}_D, \bt X}) > 0$ and $\lambda_1(S_{ \bar{\bt Y}_R \mid \bar{\bt Y}_D}) > 0$. By Lemma \ref{lemma_limit} and (\ref{eig_ubound1})
\begin{align}
\lim_{\sigma^2 \to 0} \lambda_i \leq \lim_{\sigma^2 \to 0} \frac{\lambda_{1}(S_{ \bar{\bt Y}_R \mid \bar{\bt Y}_D, \bt X})+a_1\sigma^2}{\lambda_{r - (i - 1)}(S_{ \bar{\bt Y}_R \mid \bar{\bt Y}_D, \bt X})+a'_{r - (i - 1)}\sigma^2} < \infty.
\end{align}

3) When $i > s^\mathrm{r} + t - d$, using (\ref{r'}) it is easy to see that $r' < i$, therefore, $\lambda_i(S_{ \bar{\bt Y}_R \mid \bar{\bt Y}_D}) = 0$. In this case, the $i^\mathrm{th}$ element is asymptotically reversely degraded. By Lemma \ref{lemma_limit} and (\ref{eig_ubound2})
\begin{align}
\lim_{\sigma^2 \to 0} \lambda_i \leq \lim_{\sigma^2 \to 0} \frac{a_i \sigma^2}{\lambda_{r}(S_{ \bar{\bt Y}_R \mid \bar{\bt Y}_D, \bt X})+a'_r\sigma^2} < \infty.
\end{align}
\end{IEEEproof}

\section{Proof of Theorem \ref{thm_sub_opt_dof}}
\label{app_prf_thm_sub_opt_dof}

%To calculate DoF improvement by compress-and-forward, $\Delta DoF$ defined in (\ref{dof-def}), 
\begin{IEEEproof} Fix the transmit covariance $S_{\bt X}$ to a full-rank matrix. Consider i.i.d. quantization of elements of the relay's observed vector, i.e., set $S_{\bt Q} = q\bt I_r$. To calculate DoF gain due to compress-and-forward relaying (\ref{dof-def}), we select $q$ such that the relaying capacity constraint in (\ref{opt_prob}) is satisfied with equality
\begin{align}
f_c(S_{\bt X}, \  q\bt I_r) = C_0(\rho).
\end{align}
Hence, $q$ depends on $\rho$. To characterize the asymptotic scaling of such a $q$ with $\rho$ (or equivalently with $\sigma^2$), we let
\begin{align}
x = \lim_{\sigma^2 \to 0} \frac{\log(q)}{\log(\sigma^2)},
\end{align}
and obtain
\begin{align}
x = \begin{cases}\frac{\alpha}{r'},  \ &  \alpha \leq r' \\ \frac{r+\alpha-r'}{r},  \ & \alpha \geq r'\end{cases}.
\end{align}
To see this, note that the pre-log factor of $C_0(\rho)$ is $\alpha$ in (\ref{c0}). For the pre-log factor of $f_c(S_{\bt X}, \  q\bt I_r)$, using Lemma \ref{lemma_limit} in Appendix \ref{app_lemmas}, as $\sigma^2 \to 0$ we have
\begin{align}
f_c(S_{\bt X}, \  q\bt I_r) &= \log \frac{\left|S_{ \bt Y_R \mid \bt Y_D}+q\bt I_r\right|}{\left|q\bt I_r\right|}  \nonumber \\ 
 &=  \sum_{i = 1}^{r'} \log \frac{\lambda'_{i}+a'_{i}\sigma^2+O(\sigma^4)+\sigma^{2x}}{\sigma^{2x}}  \nonumber \\
 & \qquad  +  \sum_{i = r'+1}^{r} \log \frac{a'_{i}\sigma^2+O(\sigma^4)+\sigma^{2x}}{\sigma^{2x}}.
\end{align}
Therefore, the pre-log factor of $f_c(S_{\bt X}, \  q\bt I_r)$ is
\begin{align}
\lim_{\sigma^2 \to 0} \frac{f_c(S_{\bt X}, \  q\bt I_r)}{- \log(\sigma^2)} = r'x + (r-r')(x-1)^+.
\end{align}
Solving $r'x + (r-r')(x-1)^+ = \alpha$, yields $x$.

Now, the desired DoF gain is
\begin{align}
\Delta DoF_{i.i.d.} = \lim_{\sigma^2 \to 0} \frac{I(\bt X ; \widehat{\bt Y}_R \mid \bt Y_D)}{- \log(\sigma^2)}.
\end{align}
Using Lemma \ref{lemma_limit}, as $\sigma^2 \to 0$,
\begin{align}
%\begin{multline}
&I(\bt X ; \widehat{\bt Y}_R \mid \bt Y_D)  \nonumber \\ & \qquad \qquad =  \log \frac{\left|S_{ \bt Y_R \mid \bt Y_D}+q I_r\right|}{\left|S_{ \bt Y_R \mid \bt Y_D, \bt X}+q I_r\right|} \nonumber \\ & \qquad \qquad
 = \sum_{i = 1}^{ r''} \log \frac{\lambda'_{i}+a'_i\sigma^2+O(\sigma^4)+\sigma^{2x}}{\lambda''_{i}+a''_i\sigma^2+O(\sigma^4)+\sigma^{2x}}  \nonumber \\ & \qquad \qquad \qquad
+  \sum_{i = r''+1}^{r'} \log \frac{\lambda'_{i}+a'_{i}\sigma^2+O(\sigma^4)+\sigma^{2x}}{a''_{i}\sigma^2+O(\sigma^4)+\sigma^{2x}} \nonumber \\ & \qquad \qquad \qquad \qquad
+ \sum_{i = r'+1}^{r} \log \frac{a'_{i}\sigma^2+O(\sigma^4)+\sigma^{2x}}{a''_{i}\sigma^2+O(\sigma^4)+\sigma^{2x}}  \nonumber \\ & \qquad \qquad
= \sum_{i = r''+1}^{r'} \log \frac{\lambda'_{i}+a'_{i}\sigma^2+\sigma^{2x}}{a''_{i}\sigma^2+\sigma^{2x}} + O(1).
%\end{multline}
\end{align}
Hence, the pre-log factor of $I(\bt X ; \widehat{\bt Y}_R \mid \bt Y_D)$ is
\begin{align}
\Delta DoF_{i.i.d.} = (r'-r'')\min(1,x) = (r'-r'')\min \left( 1 , \frac{\alpha}{r'} \right).
\end{align}
Finally, by comparing (\ref{det_comps}) and (\ref{dof_diff}) we have $r'-r'' = DoF_R - DoF_D$.
\end{IEEEproof}
%The last equality is (\ref{chain_dof}) in Theorem \ref{thm_finite_infinite}.

\section{Proof of Theorem \ref{thm_opt_dof}}
\label{app_prf_thm_opt_dof}%\begin{lemma} Consider $\bar{\bt Y}_r$ and $\bar{\bt Y}_d$ defined in (\ref{r_nonoise})-(\ref{d_nonoise}). Let the matrix $\widetilde{C}_r \in \mathbb{C}^{\tilde{r} \times r}$ with $\tilde{r} = \min \left(r, s, \left(r+d-t\right)^+ \right)$ be such that
\begin{IEEEproof}
Using the cut-set upper bound (\ref{csb_max_min}), we have
\begin{align}
\Delta DoF \leq \min ( DoF_R - DoF_D, \alpha).
%\label{opt_dof}
\end{align}
To almost surely achieve this upper bound by compress-and-forward, we transform $\bt Y_R$ using a matrix $\widetilde{C}_R \in \mathbb{C}^{\tilde{r} \times r}$, then describe $\widetilde{C}_R \bt Y_R$ by
\begin{align}
\widehat{\bt Y}_R = \widetilde{C}_R \bt Y_R + \bt Q,
\end{align}
with $\bt Q \sim \mathcal{CN}(\bt 0_{\tilde{r} \times 1}, q\bt I_{\tilde{r}})$ independent of other variables and
\begin{align}
\tilde{r} = \min (r, s, \left(r + d - t \right)^+ ).
\end{align}
Similar to the proof of Theorem \ref{thm_sub_opt_dof} in Appendix \ref{app_prf_thm_sub_opt_dof}, we achieve
\begin{multline}
\Delta DoF \\ = \left( \mathrm{rank}(\widetilde{C}_RS_{ \bar{\bt Y}_R \mid \bar{\bt Y}_D}\widetilde{C}_R^{\dagger})- \mathrm{rank}(\widetilde{C}_RS_{ \bar{\bt Y}_R \mid \bar{\bt Y}_D, \bt X}\widetilde{C}_R^{\dagger}) \right)  \\ \cdot \min \left( 1, \frac{\alpha}{\mathrm{rank}(\widetilde{C}_RS_{ \bar{\bt Y}_R \mid \bar{\bt Y}_D}\widetilde{C}_R^{\dagger})} \right).
\label{dof_with_com}
\end{multline}
In Theorem \ref{thm_sub_opt_dof}, without combining the relay's observed vector, we had $\mathrm{rank}(S_{ \bar{\bt Y}_R \mid \bar{\bt Y}_D, \bt X}) = r''$ and $\mathrm{rank}(S_{ \bar{\bt Y}_R \mid \bar{\bt Y}_D}) = r'$. The key step here is to design the combining matrix $\widetilde{C}_R$ at the relay such that $\mathrm{rank}(\widetilde{C}_RS_{ \bar{\bt Y}_R \mid \bar{\bt Y}_D, \bt X}\widetilde{C}_R^{\dagger}) = 0$, while $\mathrm{rank}(\widetilde{C}_RS_{ \bar{\bt Y}_R \mid \bar{\bt Y}_D}\widetilde{C}_R^{\dagger}) = r'-r''$. This can be obtained by distributed zero-forcing of interference, i.e., $\widetilde{C}_R$ is such that for some $A \in \mathbb{C}^{\tilde{r} \times d}$,
\begin{align}
\begin{bmatrix}\widetilde{C}_R \  A \end{bmatrix} \begin{bmatrix}H_{TR} \\  H_{TD} \end{bmatrix} S_{\bt X_T}^{\frac{1}{2}} = \bt 0,
\end{align}
while
\begin{align}
\mathrm{rank} \left(\begin{bmatrix}\widetilde{C}_R \  A \end{bmatrix} \begin{bmatrix}H_{SR} \\  H_{SD} \end{bmatrix} S_{\bt X}^{\frac{1}{2}} \right) = \tilde{r},
\end{align}
almost surely. In other words, $\widetilde{C}_R$ is chosen such that the observed interference at the relay is aligned with the row space of the observed interference at the destination, i.e.,
\begin{align}
\mathrm{rowspan}\left(\widetilde{C}_RH_{TR}S_{\bt X_T}^{\frac{1}{2}}\right) \subseteq \mathrm{rowspan}\left(H_{TD}S_{\bt X_T}^{\frac{1}{2}}\right).
%\begin{bmatrix} \  -A \end{bmatrix}\begin{bmatrix} H_{tr} \\ H_{td} \end{bmatrix}S_{\bt X_t}^{\frac{1}{2}} = \bt 0,
\end{align}
%\begin{align*}
%\widetilde{C}_r H_{tr}S_{\bt X_t}^{\frac{1}{2}} = -A H_{td} S_{\bt X_t}^{\frac{1}{2}}.
%\end{align*}
Note that since $\begin{bmatrix}H_{TR}^{\dagger} \  H_{TD}^{\dagger} \end{bmatrix}^{\dagger}S_{\bt X_T}^{\frac{1}{2}}$ has an $\left(r + d - t \right)^+$ dimensional left null space, such a zero-forcing matrix always exists. Now, we argue that $\mathrm{rank}(\widetilde{C}_RS_{ \bar{\bt Y}_R \mid \bar{\bt Y}_D, \bt X}\widetilde{C}_R^{\dagger}) = 0$ and $\mathrm{rank}(\widetilde{C}_RS_{ \bar{\bt Y}_R \mid \bar{\bt Y}_D}\widetilde{C}_R^{\dagger}) = r' - r''$.

We have
%\begin{align*}
%S_{ \widetilde{C}_r\bar{\bt Y}_r , \bar{\bt Y}_d} = & \left[\begin{matrix} \widetilde{C}_rH_{sr}  \\ H_{sd} \end{matrix}\right] S_{\bt X} \left[\begin{matrix} H_{sr}^\dagger\widetilde{C}_r^\dagger & H_{sd}^\dagger \end{matrix}\right] + \\ & \left[\begin{matrix} \widetilde{C}_rH_{tr} S_{\bt X_t}H_{tr}^\dagger\widetilde{C}_r^\dagger &  \widetilde{C}_rH_{tr}S_{\bt X_t}H_{td}^\dagger \\ H_{td} S_{\bt X_t}H_{tr}^\dagger\widetilde{C}_r^\dagger &  H_{td}S_{\bt X_t}H_{td}^\dagger \end{matrix}\right],
%\end{align*}
\begin{multline}
S_{ \widetilde{C}_R\bar{\bt Y}_R, \bar{\bt Y}_D}  = \left[\begin{matrix} \widetilde{C}_RH_{SR}  \\ H_{SD} \end{matrix}\right] S_{\bt X} \left[\begin{matrix} H_{SR}^\dagger\widetilde{C}_R^\dagger & H_{SD}^\dagger \end{matrix}\right]  \\ + \left[\begin{matrix} \widetilde{C}_RH_{TR}  \\ H_{TD} \end{matrix}\right] S_{\bt X_T} \left[\begin{matrix} H_{TR}^\dagger\widetilde{C}_R^\dagger & H_{TD}^\dagger \end{matrix}\right],
\end{multline}
and
\begin{align}
\mathrm{rank}(S_{ \widetilde{C}_R\bar{\bt Y}_R, \bar{\bt Y}_D}) &= \min \left( \tilde{r}+d, \ \min \left(s, \tilde{r}+d \right)+\min \left(d , t \right) \right),
\end{align}
almost surely. Similar to the proof of Lemma \ref{lemma_rank}, by rank additivity of generalized Schur complement
\begin{multline}
\mathrm{rank}\left(\widetilde{C}_RS_{\bar{\bt Y}_R \mid \bar{\bt Y}_D}\widetilde{C}_R^{\dagger}\right)  = \mathrm{rank}\left(S_{\widetilde{C}_R\bar{\bt Y}_R, \bar{\bt Y}_D}\right) - \mathrm{rank}\left(S_{\bar{\bt Y}_D}\right)  \\  = \min \left(\tilde{r}+d, d+s, s+t\right) - \min \left(d, s+t\right) =  r' - r''.
\end{multline}
Also, we have
\begin{align}
\mathrm{rank}\left(S_{ \widetilde{C}_R\bar{\bt Y}_R, \bar{\bt Y}_D, \bt X}\right)  = s^{\mathrm{r}} + \min \left( d , t \right) = \mathrm{rank}\left(S_{\bar{\bt Y}_D, \bt X}\right),
\end{align}
%\begin{align}
%&\mathrm{rank}\left(S_{ \widetilde{C}_R\bar{\bt Y}_R, \bar{\bt Y}_D, \bt X}\right) \nonumber \\ & = \mathrm{rank}\left( \left[\begin{matrix} \widetilde{C}_RH_{TR} S_{\bt X_T}H_{TR}^\dagger\widetilde{C}_R^\dagger &  \widetilde{C}_RH_{TR}S_{\bt X_T}H_{TD}^\dagger & \widetilde{C}_RH_{SR}S_{\bt X}  \\ H_{TD} S_{\bt X_T}H_{TR}^\dagger\widetilde{C}_R^\dagger &  H_{TD}S_{\bt X_T}H_{TD}^\dagger & H_{SD}S_{\bt X} \\ \bt 0 & \bt 0 & S_{\bt X} \end{matrix}\right] \right) \nonumber \\ & = s^{\mathrm{r}} + \min \left( d , t \right) \nonumber \\ & = \mathrm{rank}\left(S_{\bar{\bt Y}_D, \bt X}\right),
%\end{align}
almost surely. Similar to the proof of Lemma \ref{lemma_rank}, by rank additivity of generalized Schur complement
\begin{multline}
\mathrm{rank}\left(\widetilde{C}_RS_{\bar{\bt Y}_R \mid \bar{\bt Y}_D, \bt X}\widetilde{C}_R^{\dagger}\right)  \\ = \mathrm{rank}\left(S_{\widetilde{C}_R\bar{\bt Y}_R, \bar{\bt Y}_D, \bt X}\right) - \mathrm{rank}\left(S_{\bar{\bt Y}_D, \bt X}\right) = 0.
\end{multline}
Therefore, the DoF gain (\ref{dof_with_com}) can be written as
\begin{align}
\Delta DoF  &= \left( r'-r'' \right) \min \left( 1, \frac{\alpha}{r' - r''} \right)  \nonumber \\ &= \min  \left( DoF_R - DoF_D, \alpha \right).
\end{align}
The last equality follows by noting $r'-r'' = DoF_R - DoF_D$ from (\ref{det_comps}) and (\ref{dof_diff}).
\end{IEEEproof}

\section{Proof of Theorem \ref{thm_cg}}
\label{app_prf_thm_cg}

\begin{IEEEproof}
We first argue that a sub-optimal evaluation of compress-and-forward rate in (\ref{cf2}) is to within a constant gap of the cut-set upper bound (\ref{csb_max_min}). Then, we show that this is true for the achievable rate expression (\ref{cf1}) when optimized by Algorithm \ref{algo} as well.

Let $\bt X \sim \mathcal{CN}(\bt 0_{s \times 1}, S_{\bt X, \mathrm{CSB}}^{*})$, where $S_{\bt X, \mathrm{CSB}}^{*}$ is the global maximizer of the cut-set bound (\ref{csb_max_min}). Then, consider matrix $C_R$ that simultaneously diagonalizes $S_{ \bt Y_R \mid \bt Y_D, \bt X}$ and $S_{ \bt Y_R \mid \bt Y_D, \bt X}$, and the generalized eigenvalues $\lambda_i$'s in decreasing order as introduced in Lemma \ref{lemma_diag}. Let $\bt Q \sim \mathcal{CN}(\bt 0_{r \times 1}, S_{\bt Q, \mathrm{CG}})$, where $S_{\bt Q, \mathrm{CG}} \triangleq C_R^{-\dagger}\Sigma_{\bt Q, \mathrm{CG}} C_R^{-1}$ with diagonal $\Sigma_{\bt Q, \mathrm{CG}}$ such that the $i^\mathrm{th}$ diagonal element is
\begin{equation}
\Sigma_{\bt Q, \mathrm{CG}}^{ii} = \begin{cases} \frac{\lambda_i}{\lambda_i - 1} & \lambda_i > 1 \\  +\infty & \lambda_i = 1 \end{cases}.
\label{s_q_i_cg}
\end{equation}
Under the above input and quantization distributions, the gap between the cut-set upper bound (\ref{csb_max_min}) and the achievable rate (\ref{cf2}) is bounded as below
%\begin{multline*}
%\min \left\{  I( \bt X ; \bt Y_R, \bt Y_D) , I( \bt X ; \bt Y_D) + C_0 \right\} - \\ \min \left\{ I( \bt X ; \widehat{\bt Y}_R, \bt Y_D), \right. \\    \left. I( \bt X ; \bt Y_D) + C_0 - I( \bt Y_R ; \widehat{ \bt Y }_R \vert \bt Y_D, \bt X ) \right\}  \\ \leq \max \left\{  I( \bt X ; \bt Y_R  \mid \bt Y_D )-I( \bt X ; \widehat{\bt Y}_R  \mid \bt Y_D ) , \right. \\    \left. I( \bt Y_R ; \widehat{ \bt Y }_R \vert \bt Y_D, \bt X ) \right\} \\ = \max \left\{   \log \frac{\left|S_{ \bt Y_R \mid \bt Y_D}\right| \cdot \left|S_{\bt Q}+S_{ \bt Y_R \mid \bt Y_D, \bt X}\right|}{\left|S_{ \bt Y_R \mid \bt Y_D, \bt X}\right| \cdot \left|S_{\bt Q}+S_{ \bt Y_R \mid \bt Y_D}\right|},  \right. \\   \left. \log \frac{\left|S_{\bt Q}+S_{ \bt Y_R \mid \bt Y_D, \bt X}\right|}{\left|S_{\bt Q}\right|} \right\} \\ \stackrel{(a)}{=} \max \left\{  \sum_{i = 1}^r \log \frac{\lambda_i(\Sigma_{ \bt Q}^{ii}+1)}{\Sigma_{ \bt Q}^{ii}+\lambda_i}, \quad \sum_{i = 1}^r \log \frac{\Sigma_{ \bt Q}^{ii}+1}{\Sigma_{ \bt Q}^{ii}} \right\} \\ \leq \sum_{i = 1}^r \max \left( \log \frac{\lambda_i(\Sigma_{ \bt Q}^{ii}+1)}{\Sigma_{ \bt Q}^{ii}+\lambda_i}, \quad \log \frac{\Sigma_{ \bt Q}^{ii}+1}{\Sigma_{ \bt Q}^{ii}}\right) \\ \stackrel{(b)}{=} \sum_{i = 1}^r \log \left(2-\frac{1}{\lambda_i}\right) \stackrel{(c)}{\leq} r - (r - s)^+ = \min(r, s).
%\end{multline*}
\begin{align}
&\min \left\{  I( \bt X ; \bt Y_R, \bt Y_D) , I( \bt X ; \bt Y_D) + C_0 \right\}  \nonumber \\ & \qquad - \min \left\{ I( \bt X ; \widehat{\bt Y}_R, \bt Y_D), \right. \nonumber \\   & \quad \quad \quad \quad \quad \quad \quad  \left. I( \bt X ; \bt Y_D) + C_0 - I( \bt Y_R ; \widehat{ \bt Y }_R \vert \bt Y_D, \bt X ) \right\} \nonumber \\
& \qquad \leq \max \left\{  I( \bt X ; \bt Y_R  \mid \bt Y_D )-I( \bt X ; \widehat{\bt Y}_R  \mid \bt Y_D ) , \right. \nonumber \\   & \qquad \qquad \qquad \qquad \qquad \qquad \qquad \quad  \left. I( \bt Y_R ; \widehat{ \bt Y }_R \vert \bt Y_D, \bt X ) \right\} \nonumber \\
& \qquad = \max \left\{   \log \frac{\left|S_{ \bt Y_R \mid \bt Y_D}\right| \cdot \left|S_{\bt Q}+S_{ \bt Y_R \mid \bt Y_D, \bt X}\right|}{\left|S_{ \bt Y_R \mid \bt Y_D, \bt X}\right| \cdot \left|S_{\bt Q}+S_{ \bt Y_R \mid \bt Y_D}\right|},  \right. \nonumber \\   & \qquad \qquad \qquad \qquad \qquad \qquad \quad  \quad  \left. \log \frac{\left|S_{\bt Q}+S_{ \bt Y_R \mid \bt Y_D, \bt X}\right|}{\left|S_{\bt Q}\right|} \right\} \nonumber \\
& \qquad \stackrel{(a)}{=} \max \left\{  \sum_{i = 1}^r \log \frac{\lambda_i(\Sigma_{ \bt Q}^{ii}+1)}{\Sigma_{ \bt Q}^{ii}+\lambda_i}, \quad \sum_{i = 1}^r \log \frac{\Sigma_{ \bt Q}^{ii}+1}{\Sigma_{ \bt Q}^{ii}} \right\} \nonumber \\
& \qquad \leq \sum_{i = 1}^r \max \left( \log \frac{\lambda_i(\Sigma_{ \bt Q}^{ii}+1)}{\Sigma_{ \bt Q}^{ii}+\lambda_i}, \quad \log \frac{\Sigma_{ \bt Q}^{ii}+1}{\Sigma_{ \bt Q}^{ii}}\right) \nonumber \\ & \qquad \stackrel{(b)}{=} \sum_{i = 1}^r \log \left(2-\frac{1}{\lambda_i}\right) \stackrel{(c)}{\leq} r - (r - s)^+ = \min(r, s).
\end{align}
Here, Lemma \ref{lemma_diag} is used in equality $(a)$, equality $(b)$ follows by the choice of $\Sigma_{\bt Q}$ in (\ref{s_q_i_cg}); note that for $i \in \{ 1, \dots, r\}$, if $\lambda_i > 1$, the first and the second arguments of the maximization are, respectively, increasing and decreasing in $\Sigma_{ \bt Q}^{ii}$. Therefore, the corresponding term is minimized by equating the two and solving for $\Sigma_{ \bt Q}^{ii}$. If $\lambda_i = 1$, the corresponding term is zero by letting $\Sigma_{ \bt Q}^{ii} \to \infty$. By Theorem \ref{prop_rev_deg}, the number of such components is at least $(r - s)^+$, hence $(c)$.

Now, we argue that the achievable rate by Algorithm \ref{algo} is also to within a constant gap of the cut-set bound. Initialize Algorithm \ref{algo} with $S_{\bt X, \mathrm{CSB}}^{*}$ and update the optimal quantization noise covariance, denoted by $S_{\bt Q}^*$. By {\cite[Theorem 16.4]{nit}}, $S_{\bt Q}^*$ is the global optimum of (\ref{cf2}) at $S_{\bt X} = S_{\bt X, \mathrm{CSB}}^*$ as well. Hence, the achievable rate by Algorithm \ref{algo}, i.e., $R_{CF} = I( \bt X ; \widehat{\bt Y}_R, \bt Y_D)$ evaluated at $\left(S_{\bt X,\mathrm{CSB}}^*, S_{\bt Q}^*\right)$, is no smaller than (\ref{cf2}) evaluated at $\left(S_{\bt X, \mathrm{CSB}}^*, S_{\bt Q, \mathrm{CG}}\right)$, which is already shown to be within $\min(r, s)$ bits of the cut-set upper bound.

\end{IEEEproof}

% you can choose not to have a title for an appendix
% if you want by leaving the argument blank
%\section{}
%Appendix two text goes here.

% use section* for acknowledgment
\section*{Acknowledgment}

We thank Professor Jun Chen for informing us of \cite{JunChen}.%, after we wrote Proposition \ref{prop_opt_S_Q} using lemma \ref{lemma_diag}.

%The authors would like to thank...

% Can use something like this to put references on a page
% by themselves when using endfloat and the captionsoff option.
%\ifCLASSOPTIONcaptionsoff
%  \newpage
%\fi

% trigger a \newpage just before the given reference
% number - used to balance the columns on the last page
% adjust value as needed - may need to be readjusted if
% the document is modified later
%\IEEEtriggeratref{8}
% The "triggered" command can be changed if desired:
%\IEEEtriggercmd{\enlargethispage{-5in}}

% references section

% can use a bibliography generated by BibTeX as a .bbl file
% BibTeX documentation can be easily obtained at:
% http://mirror.ctan.org/biblio/bibtex/contrib/doc/
% The IEEEtran BibTeX style support page is at:
% http://www.michaelshell.org/tex/ieeetran/bibtex/
%\bibliographystyle{IEEEtran}
% argument is your BibTeX string definitions and bibliography database(s)
%\bibliography{IEEEabrv,../bib/paper}
%
% <OR> manually copy in the resultant .bbl file
% set second argument of \begin to the number of references
% (used to reserve space for the reference number labels box)
%\begin{thebibliography}{1}

%\bibitem{IEEEhowto:kopka}
%H.~Kopka and P.~W. Daly, \emph{A Guide to \LaTeX}, 3rd~ed.\hskip 1em plus
%  0.5em minus 0.4em\relax Harlow, England: Addison-Wesley, 1999.

%\end{thebibliography}
\bibliographystyle{IEEEtran}
\bibliography{IEEEabrv,ref_file}

% biography section
% 
% If you have an EPS/PDF photo (graphicx package needed) extra braces are
% needed around the contents of the optional argument to biography to prevent
% the LaTeX parser from getting confused when it sees the complicated
% \includegraphics command within an optional argument. (You could create
% your own custom macro containing the \includegraphics command to make things
% simpler here.)
%\begin{IEEEbiography}[{\includegraphics[width=1in,height=1.25in,clip,keepaspectratio]{mshell}}]{Michael Shell}
% or if you just want to reserve a space for a photo:

%\begin{IEEEbiography}{Michael Shell}
%Biography text here.
%\end{IEEEbiography}

% if you will not have a photo at all:
%\begin{IEEEbiographynophoto}{John Doe}
%Biography text here.
%\end{IEEEbiographynophoto}

% insert where needed to balance the two columns on the last page with
% biographies
%\newpage

%\begin{IEEEbiographynophoto}{Jane Doe}
%Biography text here.
%\end{IEEEbiographynophoto}

% You can push biographies down or up by placing
% a \vfill before or after them. The appropriate
% use of \vfill depends on what kind of text is
% on the last page and whether or not the columns
% are being equalized.

%\vfill

% Can be used to pull up biographies so that the bottom of the last one
% is flush with the other column.
%\enlargethispage{-5in}

% that's all folks
\end{document}